\documentclass[a4paper,11pt]{article}
\usepackage[geometry=standard,physics=standard,citation=biblatex,style=reiko,quiver=off]{reikoshortcut}
\usepackage{geometry}
\usepackage{textcomp}
\usepackage{amsmath}
\usepackage{amssymb}
\usepackage{graphicx}
\usepackage{color,xcolor}
\usepackage{mathrsfs}
\usepackage{caption}
\usepackage[textsize=scriptsize,textwidth=2.5cm]{todonotes}
\usepackage[colorlinks=true, allcolors=blue,  breaklinks=true]{hyperref}
\usepackage{tcolorbox}
\usepackage{tikz}
\usepackage{tikz-cd}
\usepackage{physics}
\usepackage{diagbox}
\usepackage{listings}
\usepackage{array}
\usepackage{subfig}
\usepackage{textcomp}
\usepackage[mathscr]{eucal}
\usepackage[normalem]{ulem}
\def\be{\begin{equation}}
\def\ee{\end{equation}}

\def\bea{\begin{eqnarray}}
\def\eea{\end{eqnarray}}
\def\hs{\hspace}

\newcommand{\gram}{G}

\begin{document}
\title{On Galilean Conformal Bootstrap II:  $\xi=0$ sector}
\vspace{14mm}
\author{
Bin Chen$^{1,2,3}$, Peng-xiang Hao$^4$, Reiko Liu$^1$ and Zhe-fei Yu$^3$\footnote{bchen01@pku.edu.cn, pxhao@tsinghua.edu.cn, reiko\_liu@pku.edu.cn, yuzhefei@pku.edu.cn}
}
\date{}

\maketitle

\begin{center}
	{\it
		$^{1}$School of Physics and State Key Laboratory of Nuclear Physics and Technology,\\Peking University, No.5 Yiheyuan Rd, Beijing 100871, P.~R.~China\\
		\vspace{2mm}
		$^{2}$Collaborative Innovation Center of Quantum Matter, No.5 Yiheyuan Rd, Beijing 100871, P.~R.~China\\
		$^{3}$Center for High Energy Physics, Peking University, No.5 Yiheyuan Rd, Beijing 100871, P.~R.~China\\
		$^4$Yau Mathematical Sciences Center,Tsinghua University, Beijing, 100084, China
	}
	\vspace{10mm}
\end{center}

\begin{abstract}
In this work, we continue our work on two dimensional Galilean conformal field theory (GCFT$_2$). Our previous work  (2011.11092) focused on the $\xi\neq 0$ sector, here we investigate the more subtle $\xi=0$ sector to complete the discussion. The case $\xi=0$ is  degenerate since there emerge interesting null states in a general $\xi=0$ boost multiplet. We specify these null states and work out the resulting  selection rules. Then, we compute the $\xi=0$ global GCA blocks and find that they can be written as a linear combination of several building blocks, each of which can be obtained from a $sl(2,\mathbb{R})$ Casimir equation. These building blocks allow us to give an Euclidean inversion formula as well. As a consistency check, we study  four-point functions of certain vertex operators in the BMS free scalar theory. In this case, the $\xi=0$ sector is the only allowable sector in the propagating channel. We find that the direct expansion of the 4-point function reproduces the global GCA block and is consistent with the inversion formula.

\end{abstract}

\baselineskip 18pt
\newpage

\tableofcontents{}

\newpage

\section{Introduction}
Null states in two dimensional conformal field theory is of central importance for bootstrapping exact models. A typical example is the Virasoro minimal models \cite{Belavin:1984vu}. The appearance of null states severely restricts the form of the OPEs between the primary fields $\phi_{p,q}$, which gives rise to the so-called fusion rules. When the null condition is applied to 4-point functions, it leads to  differential equations for  the 4-point function. Together with additional conditions such as the crossing symmetry, the 4-point functions and the structure constants can be determined exactly. 

The logic of higher dimensional ($d>2$) conformal bootstrap is somewhat  different. It highly depends on the convexity of the parameter space of CFT which comes from the unitarity of the theory\footnote{For a nice review on modern conformal bootstrap, see \cite{Poland:2018epd}.}. This is because in higher dimensions, the constraints from the representation theory of the conformal group, which has only finite generators, is not as strong as the ones in two dimensions.  Nevertheless, making use of the crossing symmetry as well as the  unitarity, people have succeeded in calculating numerically the conformal dimensions and OPE coefficients of various CFTs to very high accuracy, among which the most famous one is the 3D Ising  CFT \cite{Rattazzi:2008pe,ElShowk:2012ht}. What makes it possible is the observation (expectation) that certain conformal field theory  live at the boundary of theory space, which means they attain the
largest allowed value of a certain operator dimension or central charge. 

The renaissance of conformal bootstrap in the past decade has brought up huge developments in both numerical and analytic studies on CFT \cite{Komargodski:2012ek,Fitzpatrick:2012yx,Kaviraj:2015cxa,Kaviraj:2015xsa,Alday:2016njk,Caron-Huot:2017vep,Simmons-Duffin:2017nub,Mazac:2016qev,Mazac:2018mdx,Mazac:2018ycv,Mazac:2019shk,Paulos:2019gtx}, and has shed light on the AdS/CFT correspondence \cite{Heemskerk:2009pn,Penedones:2010ue,Fitzpatrick:2011ia,Alday:2016htq} and S-matrix bootstrap \cite{Paulos:2016fap,Paulos:2016but,Paulos:2017fhb,Homrich:2019cbt}.  It would be interesting to push the studies to the theories with conformal-like symmetries. 

In \cite{Chen:2020vvn} we tried to develop the bootstrap program for two dimensional Galilean conformal field theory (GCFT), based on previous works \cite{Bagchi:2016geg,Bagchi:2017cpu}. The 2D GCFT is a kind of non-relativistic conformal field theory with scaling symmetry and the
boost symmetry as follows:
\begin{equation}
\begin{aligned}
    x\rightarrow \lambda x, &\qquad y\rightarrow \lambda y,\\
x\rightarrow x,& \qquad y\rightarrow y+v x.
\end{aligned}
\end{equation}
Its symmetry algebra is called global Galilean conformal algebra and can be enhanced to the infinitely dimensional local Galilean conformal algebra (GCA). The local GCA is isomorphic to the Bondi-Metzner-Sachs (BMS) algebra in three dimensions, which generates the asymptotic symmetries of 3d 
flat spacetimes%\footnote{Due to this isomorphism, there are attempts to establish holography theory in asymptotic flat spacetimes}
. The generators of local GCA$_2$ (BMS$_3$) include the superrotations $L_n$ and the supertranslations $M_n$, satisfying the following commutation relations
\begin{equation}
\begin{aligned}
[L_n,L_m]&=(n-m)L_{n+m}+\frac{c_L}{12} n(n^2-1)\delta_{n+m,0},\\
[L_n,M_m]&=(n-m)M_{n+m}+\frac{c_M}{12} n(n^2-1)\delta_{n+m,0},\\
[M_n,M_m]&=0.
\end{aligned}
\end{equation}
The algebra is of infinite dimensions, just like the Virasoro algebra in CFT$_2$. Even though 2D Galilean conformal symmetry is infinite dimensional, it is hard to use in practice since the explicit form of local GCA block is unknown. It could be more convenient to work with global GCA blocks in bootstrapping. The global GCA block relies only on the global subalgebra\footnote{So our Galilean conformal bootstrap is more like the conformal bootstrap in higher dimensions ($d>2$).}, which is generated by $L_i, M_j, i,j=-1,0,1$. With respect to the global subalgebra, the quasi-primary operators are characterized by $\D$ and $\xi$, the eigenvalues of $L_0$ and $M_0$. One typical feature in GCA$_2$ is that the quasi-primaries generally form boost multiplets, i.e reducible but indecomposable modules. In a multiplet the quasi-primary states have the same conformal dimension and form a Jordan block under the action of $M_0$. The Hilbert space of a GCFT$_2$ is composed of the multiplets of various ranks. This is reminiscent of Logarithmic conformal field theories (LCFT), in which the quasi-primaries form dilation multiplets.  Another remarkable feature is that the states in a 2D GCFT could have negative norms, showing that the theory is generically non-unitary. This makes the study of Galilean conformal bootstrap more interesting, as it may present an example that the bootstrap could be done without unitarity.  %In this work we will follow this approach and leave the  bootstrap program based on the full GCA to the future works.}

In our previous study, we focused on the GCFT$_2$ whose spectrum includes no $\xi=0$ operators.  In \cite{Chen:2020vvn}, we discussed several important ingredients in Galilean conformal bootstrap, including the global blocks of the multiplets, the Galilean conformal partial waves (GCPWs) and the Euclidean inverse formula. We checked the robustness of the framework by studying the 4-point functions of  the Generalized Galilean Free Field Theories (GGFT), and found consistent pictures. 

In this paper we improve our framework by completing the analysis on the $\xi=0$ sector in a GCFT$_2$. As in the $\xi\neq 0$ case,  multiplets appear in the $\xi=0$ sector as well. What makes this case special is the emergence of the null states. As we introduced above, in a 2d CFT with Virasoro symmetry the null states always give strong constraints on the theory. Even though we cannot use the full power of the local GCA symmetry, the null states in the $\xi=0$ sector (which only rely on the global GCA symmetry ) do lead to some novel features. For example,  we find some selection rules for non-vanishing three point structure constant.
%\tba

The appearance of null states also has a dramatic effect on the GCA multiplet blocks. As in CFT$_2$, the null states should be modded out to get the physical Hilbert space. In a multiplet, the null states must be abandoned. The construction of the GCA multiplet block becomes subtler in the $\xi=0$ sector. We consider global GCA blocks of $\xi=0$ quasi-primaries appearing in the 4-point function of four identical external operators. Now we cannot use the same  Casimir equation as in the $\xi\neq0$ case to  determine the GCA block. Instead, we calculate the full GCA multiplet blocks directly by inserting the complete bases in the striped 4-point function. We show that the multiplet blocks can be decomposed into different $sl(2,\mathbb{R})$ blocks, each of which can be obtained from  the  Casimir equation of $sl(2,\mathbb{R})$. % By coincidence, this Casimir equation is exactly the one in 1D CFT with non-identical external operators. However,  can only be obtained by a direct calculation. \cb{}  \tba

Another tricky issue is on the harmonic analysis of the $\xi=0$ sector. In \cite{Chen:2020vvn}, we followed the study of 1D CFT \cite{Maldacena:2016hyu} and determined the Galilean conformal partial waves (GCPW) in the $\xi\neq 0$ sector. One remarkable point in the analysis is on the boundary conditions. More precisely, in the GCA case, the cross ratios $(x,y)$ reside in the region $x\in (0,2), y\in \mathbb{R}$. One need to paste  the GCPWs in $x\in (0,1)$ and $x\in (1,2)$ at $x=1$ by imposing the normalizable conditions. This strategy worked fine in the $\xi\neq 0$ sector, but it fails in the $\xi=0$ sector: the divergent terms near $x=1$ will destroy the normalizable and the hermitian condition.   In this work, we choose the alpha space approach to give an inverse formula, in order to avoid the trouble at  $x=1$. More precisely, we get an inverse formula for the building blocks, not the full multiplet blocks. In the alpha space approach, the corresponding conformal partial waves are normalizable and provides a complete orthogonal bases of the Hilbert space so that we can define an inverse function for each building block.  %}%{I modify some statements. May be we need shorten this paragraph since the inversion formula in this work is not so important (but here the introduction is a little bit long).}

Very recently,  an interesting GCFT that include a $\xi=0$ sector has been constructed \cite{Hao:2021urq}. In this so-called BMS free scalar theory, the $\xi=0$ sector appears naturally in the propagating module if we consider the 4-point functions of certain vertex operators. We investigate the decomposition of the 4-point function in this model and find a consistent picture.%  \tba

The remaining parts of this paper are organized as follows. In section 2, we give a brief review on GCFTs and Galilean conformal bootstrap in the $\xi\neq 0$ sector. In section 3, we discuss the null states in the $\xi=0$ sector and their implication on the three-point functions. In section 4, we first calculate global GCA blocks of $\xi=0$ multiplets to all orders, with the explicit forms for the first few orders, and then show that the blocks are actually composed of the $sl(2,\mathbb{R})$ blocks, which can be obtained by the Casimir equation. In section 5, we show how to get an inversion formula in the alpha space approach. In section 6, we discuss the $\xi=0$ multiplet appearing in the propagating channel in the four-point function of vertex operators in BMS free scalar theory. We end with the discussions and conclusions in section 7. %\tba

\section{Review on GCFTs}
In this section, we give a brief review on the multiplet 
and the inversion formula for the $\xi\neq 0$ sector in GCFT$_2$. For  more detailed discussions, see our previous paper \cite{Chen:2020vvn}. 

\subsection{Basics}
Two dimensional Galilean conformal field theory is a non-relativistic analogue of CFT$_2$, with the symmetry algebra $\mathcal{V}irasoro\times\mathcal{V}irasoro$ in CFT$_2$ being replaced by the 2d Galilean conformal algebra (GCA). The GCA is generated by the following commutation relations
\begin{equation}
\begin{aligned}
[L_n,L_m]&=(n-m)L_{n+m}+\frac{c_L}{12} n(n^2-1)\delta_{n+m,0} ,\\
[L_n,M_m]&=(n-m)M_{n+m}+\frac{c_M}{12} n(n^2-1)\delta_{n+m,0} ,\\
[M_n,M_m]&=0.\nn
\end{aligned}
\end{equation}
This algebra is isomorphic to BMS$_3$ and can be obtained by taking non-relativistic  (ultra-relativistic) contraction on the 2d conformal algebra.
In this paper, we mainly discuss the GCFT based on the global Galilean conformal algebra  $sl(2,\mathbb{R})\times \mathbb{R}^3$ which is the  maximal finite dimensional subalgebra. This algebra  has the following generators: 
\begin{equation*}
\begin{aligned}
      &L_{-1}: \mbox{$x$-translation}, \qquad L_0: \text{dilation}, \qquad L_1: \text{$x$-special conformal
transformation,}\\
      &M_{-1}: \text{$y$-translation}, \qquad M_0: \text{boost}, \qquad M_1: \text{$y$-special conformal
transformation.}
\end{aligned}
\end{equation*}

The operators in GCFT can be organized into  primaries and their descendants, as in CFT$_2$. The primary operators at the origin $\mathcal{O}=\mathcal{O}(0,0)$ are labeled by the eigenvalues $(\Delta, \xi)$ of $(L_0, M_0)$:
\begin{equation}
[L_0,\cO]=\Delta \cO,\qquad [M_0,\cO]=\xi \cO,
\end{equation}
where $\Delta$ and $\xi$ are referred to as the conformal weight and the boost charge  respectively. They obey the highest weight conditions
\begin{equation}
[L_n,\cO]=0,\qquad [M_n,\cO]=0, \qquad n>0.
\end{equation}
Acting $L_{-n},M_{-n}$ with $n>0$ successively on the primaries, we  get their descendants.
The operators at other positions can be obtained by the translation operator $U=e^{x L_{-1}+y M_{-1}}$,
\begin{equation}
\cO(x,y)=U\cO(0,0)U^{-1}.
\end{equation}
Using the Baker-Campbell-Hausdorff (BCH) formula, the transformation law for the primary operators are
\begin{align}
    [L_n,\cO(x,y)]&=(x^{n+1}\partial_x+(n+1)x^ny\partial_y+(n+1)(x^n\Delta+nx^{n-1}y\xi))\cO(x,y),\label{Lntrans}\\
    [M_n,\cO(x,y)]&=(x^{n+1}\partial_y+(n+1)x^n\xi)\cO(x,y),\ \ \forall n\geq-1,\label{Mntrans}
\end{align}
which can be integrated to  get the transformation law
\begin{equation}
    \cO'(x,y)=|f'|^{\D}\,e^{\xi\frac{g'+yf''}{f'}}\,\cO(x',y').  
\end{equation}
under the finite transformation $x\rightarrow f(x),y\rightarrow f'(x)y+g(x)$.

On the other hand, the operators in a  GCFT can also be organized into quasi-primary operators and their global descendant operators, which is our basic set up in this work. A quasi-primary operator transform covariantly under the global GCA symmetry. It is characterized by $(\Delta,\xi)$ as well, and its global descendant operators are constructed by acting $L_{-1}$ and $M_{-1}$ only.

By requiring the vacuum is invariant under the global symmetry, correlation functions of quasi-primary operators are well constrained. For the singlet, the two-point function
and three-point function are\cite{Bagchi:2009ca}
\begin{equation}\label{pt27}
\begin{aligned}
        G_2(x_1,x_2,y_1,y_2) &= d \,\delta_{\Delta_1,\Delta_2}\delta_{\xi_1,\xi_2}|x_{12}|^{-2\Delta_1}e^{-2\xi_1\frac{ y_{12}}{x_{12}}},\\
G_3(x_1,x_2,x_3,y_1,y_2,y_3) &= c_{123}|x_{12}|^{-\Delta_{123}}|x_{23}|^{-\Delta_{231}}|x_{31}|^{-\Delta_{312}}e^{-\xi_{123}\frac{y_{12}}{x_{12}}}e^{-\xi_{312}\frac{y_{31}}{x_{31}}}e^{-\xi_{231}\frac{y_{23}}{x_{23}}},
\end{aligned}
\end{equation}
where $d$ is the normalization factor of the two-point function, $c_{123}$ is the coefficient of three-point function which encodes dynamical information of the GCFT$_2$, and
\begin{equation}
x_{ij}\equiv x_i-x_j,\ \ y_{ij}\equiv y_i-y_j,\ \ \Delta_{ijk}\equiv\Delta_i+\Delta_j-\Delta_k,\ \ \xi_{ijk}\equiv\xi_i+\xi_j-\xi_k.
\end{equation}
The four-point functions of singlet quasi-primary operators can be determined up to an arbitrary function of the cross ratios,
\begin{equation}
G_4=\langle \prod_{i=1}^4\cO_i(x_i,y_i)\rangle =\prod_{i,j}|x_{ij}|^{\sum_{k=1}^4 -\Delta_{ijk}/3}e^{-\frac{y_{ij}}{x_{ij}}\sum_{k=1}^{4}\xi_{ijk}/3}\mathcal{G}(x,y)
\end{equation}
where the indices $i=1,2,3,4$ label the external operators $\cO_i$, $\mathcal{G}(x,y)$ is called the stripped four-point function and $x$ and $y$ are the cross ratios,
\begin{equation}
x\equiv\frac{x_{12}x_{34}}{x_{13}x_{24}}\qquad \frac{y}{x}\equiv\frac{y_{12}}{x_{12}}+\frac{y_{34}}{x_{34}}-\frac{y_{13}}{x_{13}}-\frac{y_{24}}{x_{24}}.
\end{equation}

\subsection{Multiplets}
Here we focus on the quasi-primary multiplet\footnote{Though in section \ref{example sec}, BMS free scalar indeed include primary multiplet, our attention in this paper is on the $\xi=0$ quasi-primary multiplets.}. Similar to the LCFT, the boost multiplet appear because $M_0$ acts non-diagonally on the quasi-primary states. Generically, $M_0$ acts as follows,
\begin{equation}
    [M_0,\mathbf{O}]=\boldsymbol{\xi}\mathbf{O}
\end{equation}
where $\mathbf{O}$ denotes a set of quasi-primary operators in the theory and $\boldsymbol{\xi}$ is block-diagonalized
\begin{equation}
\boldsymbol{\xi}=
\begin{pmatrix}
\ddots & & & \\
 &\boldsymbol{\xi}_i& & \\
  & & \boldsymbol{\xi}_j& \\
   & & & \ddots\\
\end{pmatrix}
\end{equation}
with
\begin{equation}
\boldsymbol{\xi}_i=
\begin{pmatrix}
 \xi_i& & & \\
  1& \xi_i& & \\
  & \ddots&\ddots &\\
   & & 1& \xi_i\\
\end{pmatrix}_{r\times r}.
\end{equation}
being the Jordan block of rank-$r$. The quasi-primaries corresponding to a rank-$r$ Jordan block form a multiplet of rank $r$. 

The two-point functions between the operators in the same multiplet can be written in the following canonical form
\begin{equation}
\langle \cO_{ia}(x_1,y_1)\cO_{jb}(x_2,y_2)\rangle =\left\{\begin{array}{ll}
0,& \mbox{for $p<0$,}\\
\delta_{ij} d_r\, |x_{12}|^{-2\Delta_i} e^{-2\xi_i\frac{y_{12}}{x_{12}}}\frac{1}{p!}\left(-\frac{2y_{12}}{x_{12}}\right)^p,& \mbox{otherwise,}\end{array} \right.
\end{equation}
where
\begin{equation}
p=a+b+1-r.
\end{equation}
In the above, the indices $i,j$ label the multiplets, $a,b$ label the $(a+1)$-th and the $(b+1)$-th operators in the multiplets $\mathcal{O}_i$ and $\mathcal{O}_j$, respectively.  One can then use the transformation rule for the multiplets,
\begin{equation}\label{mftran}
    \tilde{O}_{a}(\tilde{x}, \tilde{y})=\sum_{k=0}^{a}\frac{1}{k!}|f'|^{-\Delta}\,\partial_{\xi}^{k}e^{-\xi\frac{g'+yf''}{f'}}\,O_{a-k}(x,y),
\end{equation}
to define the out state of the quasi-primary state at infinity, which will be used to calculate the inner product and the Gram matrix,
\begin{equation}\label{outstate}
\langle O_{a}|=\lim\limits_{ y\rightarrow 0 \atop x\rightarrow \infty}\sum\limits_{k=0}^{a}\langle 0|O_{a-k}(x,y) \frac{1}{k!} \partial_{\xi}^k e^{2\xi\frac{y}{x}} x^{2\Delta}.
\end{equation}
The inner product of the quasi-primary states are
\bea 
\langle O_a|O_b\rangle &=\lim\limits_{x_1\to \infty,x_2 \to 0, \atop  y_{1}\to 0, y_{2}\to 0}\frac{1}{k!} \partial_{\xi}^k e^{2\xi\frac{y_1}{x_1}} x_1^{2\Delta} \sum_{k=0}^{a}\langle O_{a-k}(x_1,y_1) O_{b}(x_2,y_2)\rangle \nonumber\\
&= \delta_{a+b,r-1}. \label{norm}
\eea 

The general form of the three-point function is
\begin{equation}
\langle \cO_{ia}\cO_{jb}\cO_{kc}\rangle =ABC_{ijk;abc},
\end{equation}
where
\begin{equation}
\begin{aligned}
A&=\exp(-\xi_{123}\frac{y_{12}}{x_{12}}-\xi_{312}\frac{y_{31}}{x_{31}}-\xi_{231}\frac{y_{23}}{x_{23}}),\\
B&=|x_{12}|^{-\Delta_{123}}|x_{23}|^{-\Delta_{231}}|x_{31}|^{-\Delta_{312}},\\
C_{ijk;abc}&=\sum_{n_1=0}^{a-1}\sum_{n_2=0}^{b-1}\sum_{n_3=0}^{c-1}c_{ijk}^{(n_1n_2n_3)}\frac{(q_i)^{n_1}(q_j)^{n_2}(q_k)^{n_3}}{a!b!c!},\label{ABCcoef}
\end{aligned}
\end{equation}
with
\begin{equation}
q_i=\partial_{\xi_i}\ln A.
\end{equation}
Note that these expression is correct for both $\xi\neq0$ and $\xi=0$. For $\xi=0$, one should take $\xi=0$ at the end of the computation. As we need it in later sections, we write the particular case of
the three-point function of a singlet $\phi$ and a rank-2 multiplet $(\mathcal{A},\mathcal{B})^{\mathrel{T}}$ with $\xi=0$
\begin{equation}
\langle \phi(x_1,y_1)\phi(x_2,y_2)\mathcal{A}(x_0,y_0)\rangle=c_\mathcal{A}AB+c_\mathcal{B}ABC \label{3pointA},
\end{equation}
\begin{equation}
\langle \phi(x_1,y_1)\phi(x_2,y_2)\mathcal{B}(x_0,y_0)\rangle=c_\mathcal{B}AB, \label{3pointB}
\end{equation}
where $A,B,C$ can be read directly from \eqref{ABCcoef}
\begin{equation}
A=e^{-2\xi_\phi\frac{y_{12}}{x_{12}}},
\end{equation}
\begin{equation}
B=x_{12}^{\Delta-2\Delta_\phi}x_{10}^{-\Delta}x_{20}^{-\Delta},
\end{equation}
\begin{equation}
C=\frac{y_{12}}{x_{12}}-\frac{y_{10}}{x_{10}}-\frac{y_{20}}{x_{20}},
\end{equation}
and $c_\mathcal{A}, c_\mathcal{B}$ are the three-point coefficients.

For the stripped four-point function of identical external operators,  when the propagating operators have non-zero boost charge $\xi_r\neq0$, its global block expansion is
\begin{equation} \label{4pointexpansion}
\mathcal{G}(x,y)=\sum_{\cO_r,\xi_r}\frac{1}{d_r}\sum_{p=0}^{r-1}\frac{1}{p!}\sum_{a,b|a+b+p+1=r}c_ac_b\partial_{\xi_r}^pg^{(0)}_{\Delta_r,\xi_r}.
\end{equation}
where $g^{(0)}_{\Delta_r,\xi_r}$ is the global GCA block for a propagating singlet with $\xi\neq0$
\begin{equation}
    g^{(0)}_{\Delta_r,\xi_r}=2^{2\Delta_p-2}x^{\Delta_r}(1+\sqrt{1-x})^{2-2\Delta_r}e^{\frac{\xi_r y}{x\sqrt{1-x}}}(1-x)^{-1/2}.  \label{singletblock}
\end{equation}
It can be obtained from the Casimir equation. Besides, the multiplet block in \eqref{4pointexpansion} can also be obtained from the Casimir equation,  which we review next.

\subsection{Casimir equations and GCA blocks}

In this subsection, we review the GCA multiplet block for $\xi\neq 0$ by using the Casimir equations. The analysis using $C_1$ and $C_3$ is valid equally for the case $\xi\neq 0$ and $\xi=0$, but due to the presence of null states, the Casimir equations cannot lead to  the blocks in the $\xi=0$. 

For GCA$_2$, there are two Casimirs $C_1$ and $C_3=C_2^2$. and the Casimir equations take the following forms
\begin{equation}
C_1f_{\Delta,\xi}(x,y)=x^2(1-x)\partial_y^2f_{\Delta,\xi}(x,y)=\xi^2f_{\Delta,\xi}(x,y), \label{C1}
\end{equation}
\begin{equation} \label{C3}
\begin{split}
C_{3}f_{\Delta,\xi}(x,y) & ={C}_{2}^{2}f_{\Delta,\xi}(x,y)\\
 & =[(3x-2)x y\partial_{y}^{2}+2x^{2}(x-1)\partial_{x}\partial_{y}+2x^{2}\partial_{y}]^{2}f_{\Delta,\xi}(x,y)\\
 & =4\xi^{2}(\Delta-1)^{2}f_{\Delta,\xi}(x,y). 
\end{split}
\end{equation}
For the first equation \eqref{C1},  there are two independent solutions if $\xi\neq 0$:
\begin{equation} \label{sol1}
f^{(1)}_{\Delta,\xi}(x,y)=A_{\Delta,\xi}(x)e^{\frac{\xi y}{x\sqrt{1-x}}}, \qquad f^{(2)}_{\Delta,\xi}(x,y)=B_{\Delta,\xi}(x)e^{-\frac{\xi y}{x\sqrt{1-x}}},
\end{equation}
where $A_{\Delta,\xi}(x)$ and $B_{\Delta,\xi}(x)$ are general functions of $x$. Note that when $\xi=0$,  these two solutions degenerate as they become independent of $y$,
\begin{equation} \label{sol2}
f^{(1)}_{\Delta,\xi}(x,y)=f^{(2)}_{\Delta,\xi}(x,y)=A_{\Delta,\xi}(x).
\end{equation}
Actually in this case there exists another independent solution, which is a linear function of $y$
\begin{equation} \label{sol3}
f^{(3)}_{\Delta,\xi}(x,y)=C_{\Delta,\xi}(x)y.
\end{equation}

Next we solve the differential equation \eqref{C3}. For $\xi\neq 0$, after substituting  \eqref{sol1} into \eqref{C3}, we obtain the differential equations for $A_{\Delta,\xi}(x)$ and $B_{\Delta,\xi}(x)$ respectively, from which ( together with the boundary condition in the OPE limit) we get the GCA singlet block \eqref{singletblock}. For $\xi=0$, things become different:  substituting \eqref{sol2} and \eqref{sol3} into \eqref{C3},  it is easy to check \eqref{C3} is
trivially satisfied by $f^{(1)}_{\Delta,\xi}=f^{(2)}_{\Delta,\xi}$ and $f^{(3)}_{\Delta,\xi}$. In other words there is no equation to restrict $A_{\Delta,\xi}(x)$ and $C_{\Delta,\xi}(x)$.

More generally, for the block of a rank-$r$ multiplet, it satisfies the equations:
\begin{equation} \label{multiCasi}
    (C_1-\lambda_1)^rf_{\Delta,\xi,r}(x,y)=0, \qquad  (C_3-\lambda_3)^rf_{\Delta,\xi,r}(x,y)=0
\end{equation}
where $\lambda_1=\xi^2$ and $\lambda_3=4\xi^{2}(\Delta-1)^{2}$ are  the eigenvalues. For $\xi\neq0$, the block for a rank-$r$ multiplet is a linear combinations of $r$ fundamental solutions of  \eqref{multiCasi}, which are $\partial_{\xi}^sg^{(0)}_{\Delta,\xi} (s=0,1,...,r-1)$. After considering the OPE limit appropriately,  we get \eqref{4pointexpansion}. For $\xi=0$, similar to the singlet case, we can only get the solution 
\begin{equation}
   f_{\Delta,\xi,r}(x,y)=\sum_{k=0}^{2r-1}A_{\Delta,\xi,k}(x)y^k, \label{yexpansion}
\end{equation}
with $A_{\Delta,\xi,k}(x)$ being undetermined.  The above naive analysis shows that the  $\xi=0$ sector is special. This is due to the fact that there are null states in the multiplet,so we need a proper treatment to obtain the $\xi=0$ GCA block. We will discuss the global Galilean conformal block in $\xi=0$ sector in section  \ref{xi=0blocksec}. 

\subsection{GCA inversion formula} \label{secinversion}
 Here we review the GCA inversion formula for the $\xi\neq0$ case. We need the harmonic analysis based on the Casimir equations \eqref{C1} and \eqref{C3}. The goal is to find a suitable Hilbert space and its orthogonal basis, called Galilean conformal partial waves (GCPWs). 
 
 The strategy is to define the Hilbert space which makes both $C_1$ and $C_3$ Hermitian. Then the complete orthogonal basis can be obtained by solving the corresponding Casimir eigen-equations. The resulting Hilbert space is the normalizable function space $f(x,y)$ defined on a strip region\footnote{This is a result of the symmetry $1\leftrightarrow2$ or $3\leftrightarrow4$ of the four-point function.}:
 \begin{equation}
x\in[0,2],\qquad y\in(-\infty,+\infty),
\end{equation}
with the boundary conditions
\bea 
f\rightarrow 0 \text{ faster than }|y|^{-1/2}, &&\quad \text{ as }|y|\rightarrow\infty, \\
f\rightarrow 0 \text{ faster than } x^{3/2}, &&\quad \text{ as } x\rightarrow 0^+, \\
f(2,y)=f(2,-y),&&
\partial_{x}f(2,0)=0.
\eea
The inner product is defined to be
\begin{equation}
(f,g) =\int dxdy\mu(x,y)f^{*}g,\label{eq:4}
\end{equation}
where the measure can be determined by the two Strum-Liouville problems coming from the Casimir equations, 
\begin{equation}
\mu(x,y)=\frac{1}{x^4}. 
\end{equation}
Solving the Casimir eigen-equations together with the above normalizable and boundary conditions, we get the GCPW:
\begin{equation}
\Psi_{\Delta,\xi}(x,y)=
\begin{cases}
e^{\frac{i\pi}{2}(\Delta-1)}(\chi'_{\Delta,\xi}+\chi'_{\Delta,-\xi})+e^{\frac{i\pi}{2}(1-\Delta)}(\chi'_{2-\Delta,\xi}+\chi'_{2-\Delta,-\xi}),\qquad \mbox{for $1<x<2$,}\\ \\
A(\Delta)(\chi_{\Delta,\xi}+\chi_{\Delta,-\xi})+A(2-\Delta)(\chi_{2-\Delta,\xi}+\chi_{2-\Delta,-\xi}),\qquad \mbox{for $0<x<1$}.
\end{cases}
\end{equation}
where
\begin{equation}
A(\Delta)=\sin\frac{\pi\Delta}{2}+\cos\frac{\pi\Delta}{2},
\end{equation}
\begin{equation}
A(2-\Delta)=\sin\frac{\pi\Delta}{2}-\cos\frac{\pi\Delta}{2}.
\end{equation}
In GCPWs,  the quantum numbers can be in  the principal series
\begin{equation}
\Delta=1+is, \qquad s\in \mathbb{R},
\end{equation}
as well as in the discrete series
\begin{equation}
\Delta=\frac{5}{2}+2n\qquad\text{or}\qquad\Delta=-\frac{1}{2}-2n\qquad n=0,1,2,\cdots.
\end{equation}

Using the result of harmonic analysis, we can get the GCA inversion function, which is simply the inner product of the four point function and the GCPW:
\begin{equation}
    I(\Delta,\xi)=(\mathcal{G},\Psi_{\Delta,\xi}).
\end{equation}
The subtle point here is that this inversion integral is generally divergent, but since we only concern the block expansion, we can actually restrict the integral on half of the integral region. It turns out the inversion function has the following singular behaviour:
\begin{equation} \label{inversionfun1}
I(\Delta,\xi)\sim-\sum_{\Delta_m,\xi_l,k}A(2-\Delta) \Gamma(k+1)\frac{2^{2\Delta_m-2}}{(\xi-\xi_l)^{k+1}}\frac{P_{\Delta_m,\xi_l,k+1}}{\Delta-\Delta_m}+\text{shadow poles}.
\end{equation}
Note that there are un-physical shadow poles. After the contour deformation, one can read the physical spectrum
from the pole location $\Delta_m$ and $\xi_l$. The residue $P_{\Delta_m,\xi_l,k+1}$ encode the information of OPE.  Note that in the inverse function there could be generally multi-pole of order $k+1$ at $\xi=\xi_l$ in $\xi$-plane and  single pole at $\Delta=\Delta_m$ in $\Delta$-plane, which comes from the contribution of a multiplet of rank $k$ with conformal dimension $\Delta_m$ and boost charge $\xi_l$.

\section{Null states in \secmath{\xi=0} multiplets}\label{sec3}

The multiplet representations admit an indefinite inner product invariant under the Galilean conformal transformations as defined in the following way. Firstly we introduce the out-state
\begin{equation}
\bra{\op_{a}}= \lim_{\substack{q\to 0\\ x\to \inf}} |x|^{2\D} \exp(2\xi q) \bra{0} \sum_{n=0}^{a}\frac{(2q)^{n}}{n!} \op_{a-n}(x,q x),
\end{equation}
for a rank-$r$ multiplet. Inserting it into the two-point functions, we get the inner product of primary states
\begin{equation}
\braket*{\op_{a}}{\op_{b}}=\d_{a+b,r-1},
\end{equation}
which is anti-diagonal and contains $\lfloor\frac{r}{2}\rfloor$ negative norms. Denoting the descendant states at level $l=n+m$ as
\begin{equation}
\ket{a,n,m}=L_{-1}^n M_{-1}^m\ket{\op_{a}},\hspace{3ex}n,m\in \Zpositive,
\end{equation}
and using the conjugation relation $L^{\dagger}_{n}=L_{-n},\, M^{\dagger}_{n}=M_{-n}$, we get the inner product matrix 
\begin{equation}
\gram_{l}(a,b;n,m)=\bra{a,n,l-n}\ket{b,m,l-m}=\bra{\op_{a}}M_{1}^{l-n}L_{1}^{n}L_{-1}^{m}M_{-1}^{l-m}\ket{\op_{b}},
\end{equation}
which is known as the Gramian matrix, or the Shapovalov form of highest weight module \cite{humphreys2021representations}. 

Besides suffering from negative norms, the inner product can have nontrivial kernel subspace at special values of $(\D,\xi)$, and the vectors in the kernel are called null states. In this section we analyze the structure of null states by directly calculating the Gramian matrix, and discuss the implication on correlation functions.

In unitary relativistic CFTs the null states form sub-module of the highest weight module, and they could be either primary or descendants in the sub-module. Unitarity implies that we should take the maximal quotient which forces the null states to zero in the physical Hilbert space, and this leads to differential equations of correlation functions. 

{\it A priori}, in non-unitary theories like Logarithmic CFTs and the GCFTs, there is no reason to mod out the null states. However in concrete examples like the stress tensor, it turns out that we should do so. Another new phenomenon is that there are null states which is neither primary descendant nor as descendant of other primary descendant.

The null states have been analysed in relativistic CFTs, see e.g. \cite{Penedones2016,Yamazaki:2016vqi,Pasterski:2021fjn}.
For 2d Virasoro CFT, see \cite{Kac:1978ge,Feigin:1981st}, and for logarithmic multiplets of Virasoro algebra, see \cite{Hogervorst:2016itc}. For the singlets in BMS algebra, see \cite{Bagchi:2009pe}.

\subsection{An illustrated example: \secmath{\xi=0} singlets}

When $\xi\neq 0,\, \D>0$, the singlet representation, denoted as $\rep{\D,\xi}$, is irreducible and there is no null states. While if $\xi=0,\, \D>0$, the singlet representation is reducible and indecomposable, containing null states. By taking the maximal quotient, $M_{i}$ acts trivially on the remaining states, hence the action of $\isolie(2,1)$ descends to $\sllie(2,\R)$, and we get the unitary irreducible highest weight representation appearing in CFT$_{1}$, denoted as $\rep{\D}$.
\begin{table}[htbp]
    \centering
    \begin{tikzcd}
{}  \arrow[rrrrr, "L_{-1}"] \arrow[ddddd, "M_{-1}"']    &   &   &   &   &   {}\\
    & {0,0} & {1,0} \arrow[ld, "M_0"] \arrow[loop, distance=2em, in=305, out=235, red] & {2,0} & {3,0} &\\
    & {0,1} & {1,1} & {2,1} & {3,1} \arrow[u, red] \arrow[l,"L_1"']  &\\
    & {0,2} & {1,2} & {2,2} \arrow[lld,"M_1"'] \arrow[l, red]   & {3,2} &\\
    & {0,3} & {1,3} & {2,3} & {3,3} &\\
{}  &       &       &       &       &                   
\end{tikzcd}
    \caption{The action on the descendant states in singlet representation. The site $(n,m)$ stands for the descendant $\ket{n,m}$, and when $\xi=0$ the three red arrows disappear.}
    \label{tab:singletmodule}
\end{table}

This can be illustrated in Table \ref{tab:singletmodule}. Denoting the descendants $\ket{n,m}=L_{-1}^{n}M_{-1}^{m}\ket{\D,\xi}$, when $\xi=0$ it's easy to see that the states below the first row $\ket{n,m}, m\geq 1$ are null, and the descendant state $\ket{0,1}=M_{-1}\ket{\op}$ satisfies the quasi-primary conditions, hence they form a sub-representation $V_{1}$ of the singlet $\rep{\D,0}$. 

Repeating this procedure, the states $\ket{n,m}, m\geq N$ form a sub-representation $V_{N}$. We can use the Jordan-Holder composition series to characterize the nested structure of a indecomposable representation. For the $\xi=0$ singlet representation the composition series is
\begin{equation}
\dots \to V_{N} \to V_{N-1} \to \dots \to V_{1} \to V_{0}:=\rep{\D,0},
\end{equation}
and the quotient at each level is isomorphic to a $\sllie(2,\R)$ highest weight representation with weight $(\D+N)$: $V_{N}/V_{N+1}=\rep{\D+N}$.

By taking the maximal quotient, effectively we set the null states to zero, and their overlaps with the remaining physical states vanish, giving rise to differential equations of correlation functions. In the $\xi=0$ singlet example the equations are just
\begin{equation}
\pdv{y}\vev{\op_{\D,0}(\xy)\dots }=0,
\end{equation}
which means the operator $\op_{\D,0}$ is topological along the $y$-direction.

The two-point function of $\op_{\D,0}$ is the same as the one in CFT$_1$. The three-point functions containing $\op_{\D,0}$ give the fusion rules of OPE. Starting from the singlet-singlet-$\op_{\D,0}$ case, the degeneracy condition is
\begin{equation}
\pdv{y_3}\vev{\op_1 \op_2 \op_{\D,0}(\xy{3})}=0,
\end{equation}
implying that
\begin{equation}
c_{12,\xi=0}(\xi_1-\xi_2)=0,\label{fusion0}
\end{equation}
aka, either the boost charges are conserved $\xi_1=\xi_2$, or the three-point coefficient vanishes $c_{12,\xi=0}$.

% And at the level of OPE this means,
% \begin{equation}
% \op_{\D_{1},0}(x_{1})\op_{\D_{2},\xi}(\xy{2})=\sum_{\D_{3}} c_{23}\,\cD_{123}(x_{12},\pdv{y_2}) \op_{\D_{3},\xi}(\xy{2})
% \end{equation}
% where the OPE block is,
% \begin{align}
% \cD_{123}(x,\p_{y})& =x^{-\D_{123}}\sum_{m} (\frac{x\p_{y}}{2\xi})^{m} P^{(\D_{321}-1,\D_{312}-1)}_{m}(-1) \\
% & = x^{-\D_{123}} \pair(1+\frac{x\p_{y}}{2\xi})^{-\D_{132}}
% \end{align}
% in which we have used the result in section \tba.

% Notice that if $x_{12}=0$, the OPE contains no summation of descendants,
% \begin{equation}
% \op_{\D_{1},0}(x)\op_{\D_{2},\xi}(\xy)=\sum_{\D_{3}} c_{23}\, x^{-\D_{123}}  \op_{\D_{3},\xi}(\xy)
% \end{equation}
% \red{possible connection with the lightray transform.}

For the three-point functions of rank-$r_{1}$ multiplet $\op_{1}^{a}$, rank-$r_{2}$ multiplet $\op_{2}^{b}$ and $\op_{\D,0}$, we relabel the $r_1\x r_2$ number of three-point coefficients as $c^{ab}:=c_{123}^{ab1}$. One can check that if $\xi_1\neq \xi_2$, all the coefficients are forced to vanish $c^{ab}=0$. %\cb{\sout{proof by induction.}}

If $\xi_1=\xi_2$, the three-point matrix $c^{ab}$ takes the left-upper triangular form,
\begin{equation}
(c^{ab})=\bmat{c^{1}& c^{2}& c^{3} & \cdots\\c^{2}& c^{3} & \cdots& 0\\c^{3}& \cdots& & \\ \vdots & 0& & }
\end{equation}
with the rank being $\min(r_1,r_2)$.

\subsection{Gramian matrix of boost Multiplets}

Denoting the Gramian matrix of rank-$r$ multiplet at level-$l$ as
\begin{equation}
\gram_r(l,a,b)=\bmat{\bra{a,n,l-n}\ket{b,m,l-m}_r}_{(l+1)\x(l+1)},
\end{equation}
due to the derivative relation it is related to the one of singlet $\gram(l)$ by
\begin{equation}
\gram_r(l,a,b)=\frac{1}{(r-a-b+1)!}\p_{\xi}^{r-a-b+1}\gram(l).
\end{equation}
Concretely, the block matrix $\gram_r(l)_{a,b}:=\gram_r(l,a,b)$ looks like,
\begin{equation}
\gram_r(l)=\bmat{
& & \frac{1}{l!}\p_{\xi}^{l}\gram(l) & \cdots& \p_{\xi}\gram(l)& \gram(l)\\
& \iddots&\iddots & \iddots&\gram(l) & \\
\frac{1}{l!}\p_{\xi}^{l}\gram(l)&\iddots & \iddots& \iddots& & \\
\vdots&\iddots &\iddots & & & \\
\p_{\xi}\gram(l)&\iddots & & & & \\
\gram(l)& & & & & 
}.
\end{equation}

The Gramian matrix of a singlet admits a closed form
\begin{equation}
\gram_{l}(n,m)=\frac{n!m!}{(n+m-l)!}\frac{\G(2\D+l)}{\G(2\D+2l-n-m)}(2\xi)^{2l-n-m},
\end{equation}
composing a right-lower triangular matrix. Its inverse is a left-upper triangular matrix
\begin{equation}
\gram^{-1}_{l}(n,m)=\frac{(-1)^{l+n+m}}{n!m!(l-n-m)!}\frac{\G(2\D+2l-n-m-1)}{\G(2\D+l-1)}(2\xi)^{-2l+n+m}.
\end{equation}

\section{Global GCA blocks of \secmath{\xi=0} multiplets} \label{xi=0blocksec}

From the discussion in the last section, we learn that the global GCA blocks of $\xi=0$ multiplets behave quite differently from the ones of $\xi\neq0$ operators due to the special null structure in the $\xi=0$ sector. In this section, we will investigate them by directly calculating the contribution from each global descendant operators in the multiplet  by inserting the complete basis in the stripped four point function. In fact, such blocks can be decomposed into different $SL(2,R)$ blocks. We can calculate these $SL(2,R)$ blocks from the quadratic Casimir operator. The calculation here only involves the four-point functions of singlet operators with the same weight $\Delta$ and boost charge $\xi$.
%In this section, we compute the global GCA blocks of $\xi=0$ multiplets. The discussion here only concerns the 4 point function of 4 identical operators. We denote it as $\phi$, with quantum number $\Delta_\phi$, $\xi_\phi$. We also restrict the discussion here for $\phi$ to be a singlet quasi-primary.

\subsection{Direct calculation } \label{direct}

The operators in  GCFTs can be divided into the quasi-primary operators and their associated global descendant operators. Inserting the basis of the operators into a four-point function, one can get the contribution from each operator family, as the global GCA block. By calculating the contribution from the descendant operators at each level, one can obtain the GCA blocks as a power series expansion. This logic is also true for the $\xi=0$ operators, which will be studied here.

For a rank-$r$ multiplet $\mathbf{O}=(\mathcal{O}_0,...,\mathcal{O}_{r-1})^{\mathrm{T}}$, one can define the projector $|\mathbf{O}|$ onto the conformal family of $\mathbf{O}$ as
\begin{equation}
    |\mathbf{O}|\equiv \sum_{\alpha,\beta}(N^{-1})_{\alpha\beta} |\alpha\rangle\langle\beta|,
\end{equation}
where the summation of $\alpha$ and $\beta$ is over all the global descendent operators of $\mathbf{O}$, and
$N_{\alpha\beta}=\langle\alpha|\beta\rangle$ is the inner-product matrix. We should also drop all the null states in this projector. The identity operator can be decomposed into the summation over all the projectors 
\begin{equation}
    1=\sum_{\mathbf{O}}|\mathbf{O}|.
\end{equation}
Here comes the following decomposition of a stripped four-point function
\begin{equation}
\frac{\langle \phi\phi\phi\phi\rangle}{\langle \phi\phi\rangle\langle\phi\phi\rangle}=\sum_{\{\mathbf{O}\}}g_\mathbf{O}(x),
\end{equation}
where $g_\mathbf{O}(x)$ is the contribution from each operator family
\begin{equation}
g_\mathbf{O}(x)=\frac{\langle\phi\phi |\mathbf{O}|\phi\phi\rangle}{\langle \phi\phi\rangle\langle\phi\phi\rangle} \label{blockform}.
\end{equation}
Note that it contains the global GCA block together with the coefficient related to the three-point coefficients.

\subsubsection*{Global block of a $\xi=0$ singlet}
We provide a warm-up: the calculation of the global block of a $\xi=0$ singlet. Due to the analysis in section 3, it is in fact the same as a $SL(2,R)$ global block in 2d CFTs. We want to show the technique details here, which can be exploited to more complicated cases with  multiplets.

Consider a quasi-primary operator $\mathcal{O}$ with weight $\Delta$ and vanishing charge $\xi$. The global descendant states are $L_{-1}^k|\mathcal{O}\rangle$, so the corresponding projector is
\begin{equation}
|\mathcal{O}|=\sum_{k=0}^\infty\frac{L_{-1}^k|\mathcal{O}\rangle\langle \mathcal{O}|L_{1}^k}{N_k} \label{pro1},
\end{equation}
where the normalization matrix is diagonal, with diagonal elements $N_k$
\begin{equation}
N_k=\langle \mathcal{O}|L_{1}^kL_{-1}^k|\mathcal{O}\rangle.
\end{equation}
Using the commutator $[L_{1},L_{-1}]=2L_0$ $k$ times, one can move one $L_1$ to the right-hand side to touch the quasi-primary state $|\mathcal{O}\rangle$. This gives the following recursion relation
\begin{equation}
N_k=\sum_{i=0}^{k-1}2(i+\Delta)N_{k-1}=k(k-1+2\Delta)N_{k-1} \label{recursion}.
\end{equation}
With the initial value begin $N_0=1$, we have
\begin{equation}
N_{k}=\Gamma(k+1)(2\Delta)_k \label{N},
\end{equation}
where $(a)_b\equiv\frac{\Gamma(a+b)}{\Gamma(a)}$ is the raising Pochhammer symbol.

We insert the projector \eqref{pro1} into the stripped four-point function to reach the global block
%\begin{equation}
%\begin{aligned}
%    & \frac{\langle \phi(x_1)\phi(x_2)|\mathcal{O}|\phi(x_3)\phi(x_4)\rangle}{\langle \phi(x_1)\phi(x_2)\rangle\langle \phi(x_3)\phi(x_4)\rangle}\\
%     =&
%     %\sum_{k=0}^\infty\frac{\langle \phi(x_1)\phi(x_2)L_{-1}^k|\mathcal{O}\rangle\langle \mathcal{O}|L_{1}^k\phi(x_3)\phi(x_4)\rangle}{N_k\langle \phi(x_1)\phi(x_2)\rangle\langle \phi(x_3)\phi(x_4)\rangle}=
%     \sum_{k=0}^\infty\frac{\langle \phi(x_1)\phi(x_2)L_{-1}^k|\mathcal{O}\rangle}{\langle \phi(x_1)\phi(x_2)\rangle}\frac{\langle \mathcal{O}|L_{1}^k\phi(x_3)\phi(x_4)\rangle}{\langle \phi(x_3)\phi(x_4)\rangle}\frac{1}{N_k}
%     \label{block1}
%\end{aligned}
%\end{equation}
\begin{equation}\label{block1}
\frac{\langle \phi(x_1)\phi(x_2)|\mathcal{O}|\phi(x_3)\phi(x_4)\rangle}{\langle \phi(x_1)\phi(x_2)\rangle\langle \phi(x_3)\phi(x_4)\rangle}= \sum_{k=0}^\infty\frac{\langle \phi(x_1)\phi(x_2)L_{-1}^k|\mathcal{O}\rangle}{\langle \phi(x_1)\phi(x_2)\rangle}\frac{\langle \mathcal{O}|L_{1}^k\phi(x_3)\phi(x_4)\rangle}{\langle \phi(x_3)\phi(x_4)\rangle}\frac{1}{N_k}.
\end{equation}
From the discussion in section 3, we learn that there is no $y$-dependence in the block above, so that we omit all the $y$ coordinates. Next, we calculate the two factors in the summand. The  first one is
%This is just the GCA block for the operator $\mathcal{O}$. Note here that we have dropped all the $y$ coordinate in \eqref{block1} for simplicity. We can do this because in the rank 1 case, there is no $y$ dependence for the GCA block \footnote{In fact, the calculation of the GCA block for the rank 1 multiplet is the same as for the $SL(2,R)$ global block. }.
%For the first term in the summation of \eqref{block1}, the numerator is:
\begin{equation}
\langle \phi(x_1)\phi(x_2)L_{-1}^k|\mathcal{O}\rangle=\lim_{x_0\rightarrow 0}\partial_{x_0}^k\langle\phi(x_1)\phi(x_2)\mathcal{O}(x_0)\rangle.
\end{equation}
%it can also be written as:
%\begin{equation}
%\langle \phi(x_1)\phi(x_2)L_{-1}^k|\mathcal{O}\rangle=\lim_{x_0\rightarrow 0}\partial_{x_0}^k\langle\phi(x_1)\phi(x_2)\mathcal{O}(x_0)\rangle \label{3-pt1}
%\end{equation}
Based on the correlation functions \eqref{pt27}, we arrive at
\begin{equation}
\frac{\langle \phi(\infty)\phi(1)L_{-1}^k|\mathcal{O}\rangle}{\langle \phi(\infty)\phi(1)\rangle}=c_{\mathcal{O}}(-1)^k(-\Delta,k)\Gamma(k+1),
\end{equation}
Here we have used the symmetry to put the four positions to $(\infty,1,x,0)$ where $x$ is the invariant cross ratio.
%get the dependence on the cross ratio.%
Additionally, $(a,b)\equiv \frac{\Gamma(a+1)}{\Gamma(b+1)\Gamma(a-b+1)}$ is the binomial coefficient, and $c_{\mathcal{O}}$ is the three-point coefficient in $\langle\phi\phi\mathcal{O}\rangle$.
Similarly, the second factor involves
\begin{equation}
\langle \mathcal{O}|L_1^k \phi(x_3)\phi(x_4)\rangle=\lim_{x_0\rightarrow\infty}x_0^{2\Delta}(-x_0^2\partial_{x_0}-2x_0\Delta)^k\langle \mathcal{O}(x_0)\phi(x_3)\phi(x_4)\rangle.
\end{equation}
One can deal with it in the inversion coordinate $w=\frac{1}{x}$, as
\begin{equation}
\frac{\langle \mathcal{O}|L_1^k \phi(x)\phi(0)\rangle}{\langle \phi(x)\phi(0)\rangle}=\frac{\langle \phi(\infty)\phi(\frac{1}{x})L_{-1}^k|\mathcal{O}\rangle}{\langle \phi(\infty)\phi(\frac{1}{x})\rangle}=c_{\mathcal{O}}(-1)^k(-\Delta,k)\Gamma(k+1)x^{\Delta+k}.
\end{equation}
Altogether, we can express the contribution from a $\xi=0$ singlet as
\begin{equation}
g_{\mathcal{O}}(x)=\sum_{k=0}^{\infty}c_O^2\frac{x^{\Delta+k}(-\Delta,k)^2\Gamma(2\Delta)\Gamma(k+1)}{\Gamma(2\Delta+k)}=c_{\mathcal{O}}^2x^{\Delta}\ _2F_1(\Delta,\Delta,2\Delta,x).
\end{equation}
There is exactly the $SL(2,R)$ block, as expected.

\subsubsection*{Global block of a rank-2, $\xi=0$ multiplet}
Considering a rank-2 multiplet $\mathbf{\mathcal{O}}=(\mathcal{O}_0,\mathcal{O}_1)^\mathrm{T}$
\begin{equation}
L_0\mathcal{O}_0=\Delta\mathcal{O}_0,\ \ \ L_0\mathcal{O}_1=\Delta\mathcal{O}_1,
\end{equation}
\begin{equation}
M_0\mathcal{O}_0=0,\ \ \ M_0\mathcal{O}_1=\mathcal{O}_0,
\end{equation}
the global descendant states are in general
\begin{equation}
L_{-1}^aM_{-1}^b|\mathcal{O}_0\rangle,\ \ \ L_{-1}^cM_{-1}^d|\mathcal{O}_1\rangle. \label{dec1}
\end{equation}
The fact that the null condition has the following form
\begin{equation}
M_{-1}\mathcal{O}_0=M_{-1}^2\mathcal{O}_1=0,
\end{equation}
\begin{equation}
L_{-1}\mathcal{O}_0=\Delta M_{-1}\mathcal{O}_0,
\end{equation}
helps us restrict our attention to the following global descendant states
\begin{equation}
A_k=L_{-1}^k|\mathcal{O}_1\rangle,\ \ \ k\geq1,
\end{equation}
\begin{equation}
B_k=L_{-1}^k|\mathcal{O}_0\rangle,\ \ \ k\geq1.
\end{equation}
From the conjugate and inner product discussed in section 2, we find the inner product at level $k$ as follows,
\begin{equation}
\langle \mathcal{A}_k|\mathcal{A}_k\rangle=0,\ \ \langle \mathcal{A}_k|\mathcal{B}_k \rangle=N_k,\ \ \langle \mathcal{B}_k|\mathcal{B}_k\rangle=0,  \label{norm}
\end{equation}
The projector of the multiplet $\mathbf{\mathcal{O}}$ can be written as
\begin{equation}
|\mathbf{\mathcal{O}}|=\sum_{k=0}^{\infty}\sum_{i,j}(P_{ij}^k)(N_k^{-1})_{ij}, \qquad i,j=1,2 \label{projector},
\end{equation}
where
\begin{equation}
P_{11}^k=|\mathcal{A}_k\rangle\langle \mathcal{A}_k|,\ \ P_{12}^k=|\mathcal{A}_k\rangle\langle \mathcal{B}_k|,\ \ P_{22}^k=|\mathcal{B}_k\rangle\langle \mathcal{B}_k|.
\end{equation}
$N^{-1}_{k}$ is the inverse of the normalization matrix given by \eqref{norm}, under the identification $1\Longleftrightarrow \mathcal{A}_k$, $2\Longleftrightarrow \mathcal{B}_k$.
Inserting the projector \eqref{projector} into the stripped four-point function, similar with the calculation in the singlet case, we can express the contribution from a rank-2 multiplet as
\begin{equation}
g_{\mathbf{O}}=\sum_{k=0}^\infty\frac{f_\mathcal{A}(1/x,-y/x^2)f_\mathcal{B}(1,0)+f_\mathcal{B}(1/x,-y/x^2)f_\mathcal{A}(1,0)}{N_k} \label{blockrank2},
\end{equation}
where
\begin{equation}
f_\mathcal{A}(x,y)=\frac{\langle \phi(\infty,0)\phi(x,y)L_{-1}^k|\mathcal{A}_k\rangle}{\langle \phi(\infty,0)\phi(x,y)\rangle}, \qquad
f_\mathcal{B}(x,y)=\frac{\langle \phi(\infty,0)\phi(x,y)L_{-1}^k|\mathcal{B}_k\rangle}{\langle \phi(\infty,0)\phi(x,y)\rangle}.
\end{equation}
Using the form of the 3-point function \eqref{3pointA} and \eqref{3pointB}, we have
\begin{equation}
f_\mathcal{A}(x,y)=(-1)^k\Gamma(k+1)x^{-1-\Delta-k}(c_1x(-\Delta,k)+c_0y(-1-\Delta,k)),
\end{equation}
\begin{equation}
f_\mathcal{B}(x,y)=(-1)^k\Gamma(k+1)x^{-\Delta-k}c_0(-\Delta,k).
\end{equation}
Substituting them back into \eqref{blockrank2}, we obtain the contribution of a rank-2 $\xi=0$ multiplet
\begin{equation}
g_{\mathbf{O}}=2c_0c_1x^\Delta\ _2F_1(\Delta,\Delta,2\Delta;x)+c_0^2x^\Delta\frac{y}{x}\ _2F_1(\Delta,1+\Delta,2\Delta;x).
\end{equation}
For simplicity, we can denote it as:
\begin{equation}
g_{\mathbf{O}}=A_0g_0+A_1g_1
\end{equation}
where
\bea 
A_0=2c_0c_1, &&g_0=x^\Delta\ _2F_1(\Delta,\Delta,2\Delta;x),\\
A_1=c_0^2, &&g_1=x^\Delta\frac{y}{x}\ _2F_1(\Delta,1+\Delta,2\Delta;x).
\eea 

\subsection*{Global block of a rank-$r$, $\xi=0$ multiplet}
In the above discussions, we have learnt the framework and techniques of direct calculation, especially the calculation of the three-point functions in the two factors as well as the inverse Gram matrix prescription for the multiplet. Now we can consider the contribution from a generic rank-$r$, $\xi=0$ multiplet quasi-primary operator.

We first calculate the two three-point functions as follows. Recall that the general three-point function of a singlet, a singlet and a rank-$r$ multiplet is
\begin{equation}
\langle \phi(x_1,y_1)\phi(x_2,y_2)O_i(x_0,y_0)\rangle=\sum_{k=0}^ic_kABC^k
\end{equation}
where $i=0,\cdots,r-1$, and $O_i$ is one of the components of the multiplet. Denote
\begin{equation}
f_{k,a,b}(x,y)=\frac{\langle \phi(\infty,0)\phi(x,y)L_{-1}^{k-a}M_{-1}^a|O_b\rangle}{\langle \phi(\infty,0)\phi(x,y)\rangle}
\end{equation}
where $L_{-1}^{k-a}M_{-1}^a|O_b\rangle$ is a global descendant operator at level $k$, related to the quasi-primary component $O_b$.
The $a=0$ case involves no partial derivative with respect to $y$, which is much easier to deal with. After taking the limit to infinity, one gets
\begin{equation}
f_{k,0,b}(x,y)=(-1)^k\Gamma(k+1)x^{-\Delta-k}\sum_{i=0}^{b}c_{b-i}\frac{q^i(-\Delta-i,k)}{i!}.
\end{equation}
For the general $a$ case, one can take the $y$-derivative first and put the two $y$-position to zero, then deal with the $x$-part. We arrive at
\begin{equation}
f_{k,a,b}(x,y)=(-1)^kx^{-\Delta-k}\sum_{i=0}^{b}c_{b-i}q^{i-a}\frac{\Gamma(1-\Delta-i)}{\Gamma(1+a-\Delta-i-k)\Gamma(1-a+i)}.
\end{equation}
where we denote
\begin{equation}
q=\frac{y}{x}.
\end{equation}
for simplicity. Then we should calculate the inverse Gram matrix after modding out all the null states. Denote
\begin{equation}
|k,a,b\rangle=L_{-1}^{k-a}M_{-1}^{a}|O_b\rangle,\ \ b=0,\cdots
,r-1.
\end{equation}
The basis at level $k$ is
\begin{equation} \label{xi=0basis}
|k,a,b\rangle,\ \ a\le[b/2],\ \ b=0,\cdots
,r-1.
\end{equation}
We can find the numbers of these operators for a rank-$r$ multiplet. It reads
\begin{equation}
\frac{r(r+2)}{4},\hs{3ex} \mbox{$r$ is even},
\end{equation}
\begin{equation}
\frac{(r+1)^2}{4},\hs{3ex} \mbox{$r$ is odd.}
\end{equation}
From the discussion in section 3, we first calculate the inner product among these states
\begin{equation}
g(k,a_1,a_2,b_1,b_2,r)=\langle k,a_1,b_1|k,a_2,b_2\rangle.
\end{equation} 
The inner product vanishes if the two states are at different levels. In this notation,
\begin{equation}
g(k,a_1,a_2,b_1,b_2,r)=\frac{1}{(-r+b_1+b_2+1)!}\partial_{\xi}^{-r+b_1+b_2+1}g(k,a_1,a_2)
\end{equation}
where
\begin{equation}
g(k,a_1,a_2)=\frac{(2\xi)^{a_1+a_2}(k-a_1)!(k-a_2)!\Gamma[2\Delta+k]}{(k-a_1-a_2)!\Gamma[a_1+a_2+2\Delta]}.
\end{equation}
After some detailed calculations, one can find that the inner product vanishes if $p\neq a_1+a_2$, where
\begin{equation}
p=b_1+b_2-r+1.
\end{equation}
The inner product depends on $k,p,a_1,a_2$ only. Since we focus on the same $k$, we omit $k$ for simplicity. In this way, we can denote the inner product as
\begin{equation}
g(k,a_1,a_2,b_1,b_2,r)=w(p,a_1,a_2).
\end{equation}

To go further, we can re-arrange the order of the bases to make the structure manifest as follows. We first organize the bases according to $c=b-a$, $c\in[0,r-1]$, then put them in order of $a\in[0,\mbox{Min}(r-1-c,c)]$. For example, consider the four bases in the rank-3 case, $|k,a,b\rangle$, where $(a,b)=(0,0),(0,1),(0,2),(1,2)$. The order is the following,
\bea
c=0,&& (a,b)=(0,0),\nn\\
c=1,&& (a,b)=(0,1),(1,2),\nn\\
c=2,&& (a,b)=(0,2).
\eea
Here comes two results in order. The first one is that the inner product of two basis states with $c_1+c_2\neq r-1$ vanishes. The other one is that the Gram matrix is block-diagonalized in the off-diagonal line. Now the question reduces to computing the inverse of the following matrix,
\begin{equation}
F_{mn}(a)=w(m+n,m,n),\ \ m,n=0,\cdots,a,\hs{3ex}a\le [(r-1)/2].
\end{equation}
The matrix $F_{mn}(a)$ is the inner product of the states with $c_1+c_2=r-1$. Also it depends on $a_1,a_2$ only, where we denote them as $m,n$. Besides, there are at most $[\frac{r-1}{2}]$ states involved in the rank-$r$ case. The inner product depends on the number of the states with the same $c$, where we denote is as $a$. We find that the inverse matrix has the following form,
\begin{equation}
(F^{-1})_{mn}(a)=\sum_{i=0}^{a}z(i,a,m,n),
\end{equation}
where 
\begin{equation}
\begin{aligned}
  &z(i,a,m,n)=\frac{(-1)^{a+m+n}}{2^{m+n}(i+k-m-n)}\\
  &\times\frac{\Gamma(2\Delta+m+n-1)\Gamma(2\Delta+m+n+a-i)}{\Gamma[i+1]\Gamma[k-a]\Gamma[m+i-1]\Gamma[a+i+1-m-n]\Gamma[n+i-1]\Gamma[2\Delta+k+a]\Gamma[2\Delta+m+n-1-i]}.
\end{aligned}
\end{equation}
This can be checked directly by the expression of the Gram matrix and the relation to $F_{mn}(a)$. Collecting all the results above, one finally gets the contribution from a rank-$r$ multiplet as
\begin{equation}
g=\sum_{k=0}^{\infty}\sum_{c=0}^{r-1}\sum_{m=0}^t\sum_{n=0}^t F_{mn}^{-1}(t)f_{m,m+c}\bar{f}_{n,r-1-c+n},
\end{equation}
where
\begin{equation}
t=\mbox{Min}(c,r-1-c),\ f_{a,b}=f_{k,a,b}(\frac{1}{x},-\frac{y}{x^2}),\ \ \bar{f}_{a,b}=f_{k,a,b}(1,0).
\end{equation}
Before going to the general result, we consider a special case with the three-point coefficients $c_i=0,\ \ i\neq0$. In this case, we have
\begin{equation}
g=\frac{1}{\Gamma[r]}x^{\Delta} q^{r-1}c_0^2\ _2F_1(\Delta,\Delta+r-1,2\Delta,x).
\end{equation}
In general, the contribution can be expressed as
\begin{equation} \label{xi=0block1}
g=\sum_{i,j|i+j+p=r-1}c_ic_jg_{ij}=\sum_{p=0}^{r-1}\sum_{i=0}^{r-1-p}c_ic_jg_{ij}|_{j=r-1-p-i},
\end{equation}
where
\begin{equation} \label{xi=0block2}
g_{ij}=\frac{1}{p!}x^\Delta q^p \sum_{a=0}^{\min(j,r-1-j)}x^af_a\ _2F_1(\Delta+a,\Delta+a+p,2\Delta+2a,x).
\end{equation}
The coefficients $f_a$ can be written into a double summation,
\begin{equation}f_{a}=\frac{\sin (\pi  \Delta ) \Gamma
   (a+2 \Delta -1) \sin (\pi
   (\Delta +p)) \Gamma (a+p+\Delta
   )}{\pi ^{3/2} \Gamma (a+1)
   \Gamma \left(a+\Delta
   -\frac{1}{2}\right)}\sum_{b,c=0}^a
f_{bc}^{(1)}f_{bc}^{(2)},
\end{equation}
where
\begin{equation}
f_{bc}^{(1)}=\frac{(-1)^{-a+b+c} \Gamma
   (-c-\Delta +1) 2^{-2 a-b-c-2
   \Delta +2} \Gamma (b+c+2 \Delta
   -1) \Gamma (-b-p-\Delta
   +1)}{\Gamma (b+c+1)},
\end{equation}
and
\begin{equation}
f_{bc}^{(2)}=
\, _4\tilde{F}_3(1,-a,-b-c,a+2
   \Delta -1;1-b,1-c,2 \Delta -1;1).
\end{equation}
We list the first few order terms explicitly 
\begin{equation}
f_0=1,\ \ f_1=-2\Delta-p,\ \ f_2=\frac{(3+8\Delta)(3+2p(2+p)+11\Delta+9p\Delta+8\Delta^2)}{16+32\Delta}.
\end{equation}
In the discussion above, we organize the contribution in terms of $g_{ij}$, with the coefficients $c_ic_j$. However, in practice, it is more convenient to organize them in terms of $q^p\tilde{F}(a,p)$,
\begin{equation}
\tilde{F}(a,p)=\frac{1}{p!}x^{\Delta+a} \ _2F_1(\Delta+a,\Delta+a+p,2\Delta+2a,x).
\end{equation}
As we will see in the next subsection, this corresponds to decompose the global $\xi=0$ block into different $SL(2,R)$ blocks. In this way, one can collect the contribution with different $c_ic_j$ into the same $\tilde{F}(a,p)$,
\begin{equation}
g=\sum_{a=0}^{[\frac{r-1}{2}]}\sum_{j=a}^{r-1-a}\sum_{i=0}^{r-1-j}c_ic_jf_aq^{r-1-i-j}\tilde{F}(a,r-1-i-j).
\end{equation}
We can further express it as
\begin{equation}\label{block462}
g=\sum_{p=0}^{r-1}q^p\sum_{a=0}^{[\frac{r-1}{2}]}\sum_{j=a}^{r-1-a}c_ic_jf_a\tilde{F}(a,p)|_{i=r-1-p-j},
\end{equation}
which turns out to be useful to read data in practice.

\subsection{Building blocks from the Casimir equation} \label{Casimirmethod}

One usual way to calculate the global blocks in GCFTs and 2d CFTs is solving the eigenequations of the Casimir operators of the corresponding algebra. This method is also helpful in the discussion of the harmonic analysis, where one wants to make the Casimir operators hermitian so that one can expand  4-point functions in terms of the eigenfunctions. In section 2, we have reviewed how this procedure goes well for  $\xi\neq0$ operators in GCFTs but fails for the $\xi=0$ sector. As shown in section 2, the Casimir equations \eqref{C1} and \eqref{C3} cannot determine the global block for the $\xi=0$ multiplet completely, since the solutions to \eqref{C1} satisfy the second Casimir equation \eqref{C3} automatically. We need to find another way to organize the global descendant states in  certain $\xi=0$ multiplet module. In fact, the GCA global block in $\xi=0$ sector can actually be written in terms of $SL(2,R)$ global blocks, at least in the case where the external operators are identical singlets which we focus on in this paper.

From the discussion in section \ref{sec3} and section \ref{direct}, we know that when $\xi=0$, the real bases of the descendant states are of the form \eqref{xi=0basis}. At each level, there exists at least one $SL(2,R)$ quasi-primary state as the linear combination of \eqref{xi=0basis}. One can further show that each state in the $\xi=0$ multiplet module is either a $SL(2,R)$ quasi-primary state, or a descendant state generated by $L_{-1}$ acting on the $SL(2,R)$ quasi-primary states. So the $\xi=0$ multiplet module $V_{\Delta,\xi=0}$ can be decomposed into different $SL(2,R)$ modules $V'_{\Delta'}$ with weight $\Delta'=\Delta+a$, where $a\leq [\frac{r-1}{2}]$,
\begin{equation}
V_{\Delta,\xi=0}=\sum_{\Delta'}\oplus V'_{\Delta'},\ \ \Delta'\in[\Delta,\Delta+[\frac{r-1}{2}]].
\end{equation}
The above fact suggests that it is feasible to use the $SL(2,R)$ blocks to build  the $\xi=0$ global blocks. The $SL(2,R)$ blocks  can  be calculated from the eigenfunction of the quadratic Casimir operator of $SL(2,R)$,
\begin{equation}
    \mathcal{C}_2=L_0^2-\frac{1}{2}\{L_{-1},L_1\}.
\end{equation}
Due to the fact that this Casimir operator commutates with $L_{-1}$, one has the following equation,
\begin{equation}
\mathcal{C}_2V'_{\Delta'}=\Delta'(\Delta'-1)V'_{\Delta'},
\end{equation}
where the module $V'_{\Delta'}$ has collected the contribution of a $SL(2,R)$ quasi-primary state with weight $\Delta'$ and all its descendant states generated by $L_{-1}$. From the projector \eqref{pro1}, one finds
\begin{equation}
    \mathcal{C}_2|\mathcal{O}|=|\mathcal{O}|\mathcal{C}_2=\lambda_\Delta|\mathcal{O}|,
\end{equation}
where $\mathcal{O}$ here refers to the $SL(2,R)$ quasi-primary operators. The analysis in this way is fine but not good enough to see the multiplet structure. 

To see more precisely the contribution from individual $SL(2,R)$ blocks, one can furthermore  consider the action of the Casimir operator on the four-point functions,
\begin{equation}
    \mathcal{D}_{1+2}\hat{g}_{\mathcal{O}}(x_i,y_i)=\lambda_\Delta\hat{g}_{\mathcal{O}}(x_i,y_i), \qquad i=1,2,3,4,     \label{twoparticle}
\end{equation}
where $\mathcal{D}_{1+2}$ is the action of the Casimir $\mathcal{C}_2$ on the first two operators and $\hat{g}_{\mathcal{O}}$ is related to the block $g_{\mathcal{O}}$ by
\begin{equation}
    \hat{g}_{\mathcal{O}}(x_i,y_i)=\langle\phi(x_1)\phi(x_2)\rangle\langle\phi(x_1)\phi(x_2)\rangle g_{\mathcal{O}}(x_i,y_i)\equiv f(x_i,y_i)g_{\mathcal{O}}(x_i,y_i) \qquad i=1,2,3,4, \label{kine}
\end{equation}
where $f(x_i,y_i)$ is a kinematic factor. 
Note that \eqref{twoparticle} is a standard eigen-equation but not in the typical form $(\mathcal{D}-\lambda_\mathcal{D})^rf=0$ for a rank-$r$ multiplet, since we are consider the building blocks instead of the entire $\xi=0$ block. Different from 2D CFTs, the action $\mathcal{D}_{1+2}$ has $y$ dependence here, recalling the action of the $SL(2)$ generators in the GCFT 
\begin{equation}
    L_{-1}=\partial_x, \quad L_0=x\partial_x+\Delta+y\partial_y, \quad L_1=x^2\partial_x+2\Delta x+2xy\partial_y+2y\boldsymbol{\xi} \label{generator}.
\end{equation}
The action of $L_1$ is not diagonalized, since there is a $\boldsymbol{\xi}$ term. As usual, the two particle realization $\mathcal{D}_{1+2}$ is
\begin{equation}
\begin{aligned}
    \mathcal{D}_{1+2}&=(L_{0}^{(1)}+L_{0}^{(2)})^2-\frac{1}{2}\{(L_{-1}^{(1)}+L_{-1}^{(2)}),(L_{1}^{(1)}+L_{1}^{(2)})\}\\
    &=\mathcal{C}_2^{(1)}+\mathcal{C}_2^{(2)}+2L_{0}^{(1)}L_{0}^{(2)}-L_{-1}^{(1)}L_{1}^{(2)}-L_{1}^{(1)}L_{-1}^{(2)}  \label{D1+2}
\end{aligned}
\end{equation}
where the  superscripts $(1)$ and $(2)$ denote the action on $\phi(x_1,y_1)$ and $\phi(x_2,y_2)$ respectively, and  $\mathcal{C}_2^{(1)}$ and $\mathcal{C}_2^{(2)}$ are the one-particle realizations of the Casimir. One can remove the kinematic factor $f(x_i,y_i)$ in \eqref{kine} and then  find that the building blocks obey the following differential equation
\begin{equation}
    \mathcal{D}g(x,y)=\lambda_\Delta'\mathcal{D}g(x,y).
\end{equation}
where
\begin{equation}
\mathcal{D}=x^2(1-x)\partial_x^2-x^2q\partial_x\partial_q-x^2\partial_x ,\ \ \ \lambda_{\Delta'}=\Delta'(\Delta'-1). \label{Dform}
\end{equation}
In practice, in order to remove the kinematic factor $f(x_i,y_i)$ in \eqref{kine}  to obtain the operator $\mathcal{D}$, one can use (\ref{generator}, \ref{D1+2}) and set the quantum numbers to zero: $\Delta=\Delta_\phi=0$ and $\xi=\xi_\phi=0$ (then the kinematic factor $f(x_i,y_i)=1$). Note that unlike the 1D case, the contribution of the one-particle realization part $\mathcal{C}_2^{(1)}$ and $\mathcal{C}_2^{(2)}$ are not zero. For the singlet, the one-particle quadratic Casimir is of  the form
\begin{equation}
    \mathcal{C}_2=\Delta^2-\Delta+2\Delta y\partial_y+2\xi y\partial_x+y^2\partial_y^2.
\end{equation}
After taking the quantum numbers to be zero, the term $y^2\partial_y^2$ still survives and will contribute to $\mathcal{D}$.  In contrast,  in 1D CFT case, there is no $y$-dependent term in $\mathcal{C}_2$ so that the contributions of $\mathcal{C}_2^{(1)}$ and $\mathcal{C}_2^{(2)}$ simply vanish.  In the conformal frame $\{(x_1,y_1),(x_2,y_2),(x_3,y_3),(x_4,y_4)\}=\{(0,0),(x,y),(1,0),(\infty,0)\}$, we find the explicit form \eqref{Dform} of $\mathcal{D}$ in terms of cross-ratios. In the following calculations and discussions, it is more convenient to use the coordinate $q=\frac{y}{x}$  instead of the coordinate $y$.

As in the $\xi\neq 0$ case,  the global block of a rank-$r$ multiplet in $\xi=0$ sector is a polynomial of degree at most $r-1$ in $q$,
\begin{equation}
g_{\mathbf{O}}(x,q)=\sum_{k=0}^{r-1}q^k\sum_{a=0}^{[\frac{r-1}{2}]}C_{k,a}w_{k,\Delta+a}(x) \label{qexpan}.
\end{equation}
Here $C_{k,a}$'s encode dynamical information, though its relation to the OPE coefficients is complicated.
The expansion in terms of $q$ is feasible, due to the fact that the action of $\mathcal{D}$ keeps the powers of $q$ in each term. As a result, the above differential equation  reduces to the eigen-equations for every individual $w_{k,\Delta+a}(x)$,
\begin{equation}
\mathcal{D}_kw_{k,\Delta'}(x)=\Delta'(\Delta'-1)w_{k,\Delta'}(x)  \label{wequ},
\end{equation}
where
\begin{equation}
\mathcal{D}_k=x^2(1-x)\partial_x^2-(1+k)x^2\partial_x. \label{D_k}
\end{equation}
Note that for a singlet with $k=0$, this is nothing but the Casimir differential operator for $SL(2,R)$, as in 2d CFTs.
The two solutions of \eqref{wequ} are respectively,
\bea
w_{k,\Delta'}^{(1)}(x)&=&x^{\Delta'}\ _2F_1(\Delta',\Delta'+k,2\Delta';x),  \label{s1}\\
w_{k,\Delta'}^{(2)}(x)&=&x^{1-\Delta'}\ _2F_1(1-\Delta',1-\Delta'+k,2-2\Delta';x).  \label{s2}
\eea
From the monodromy condition, we know $w^{(1)}_{k,\Delta'}$ is the conformal block itself, while $w^{(2)}_{k,\Delta'}$ is the shadow block. These eigenfunctions provide the building blocks for the global GCA blocks in $\xi=0$ sector,  just the ones shown by direct calculation in \ref{direct}.

\section{Inversion formula} \label{inversionsec}
%\yu{}{This section is revised to include the most part of the previous appendix A. So I delete  the appendix A.}

We have reviewed the GCA inversion formula in \ref{secinversion}, which is derived under the assumption that no  $\xi=0$ operator appears in the theory. In this section, we try to give an inversion formula, including  operators with $\xi=0$.  Our inversion formula is based on the alpha space approach for the building blocks \eqref{s1}, not the full $\xi=0$ conformal blocks \eqref{xi=0block2} .

\subsection{Basic logic}

We have seen that due to the emergent null states, the $\xi=0$ block is not simply the  analytic continuation of the $\xi\neq0$ block.  For the $\xi=0$ sector, we need a different decomposition of the four-point function in terms of Galilean conformal partial waves.

As the first step, we need to use the inversion integral in section \ref{secinversion} to obtain the inversion function $I(\Delta,\xi)$. For the propagating operators with $\xi\neq0$, we just use the inversion formula in section \ref{secinversion}. If there are propagating operators  with $\xi=0$, then the inversion function will have (multi) pole at $\xi=0$. Reading the corresponding residue in the inversion function gives us the contribution $\mathcal{G}^0$ of all  $\xi=0$ propagating operators in the 4-point function.   %\footnote{Which has a same form as   in \ref{4pointexpansion}, but with $\xi=0$. This is of  course not the $\xi=0$ GCA block expansion.}, where  $\mathcal{G}^0$ is .}

However, the $\xi=0$ sector is special. In section  \ref{Casimirmethod} we show that the Casimir method can only derive the building blocks  \eqref{s1}, not the full multiplet block \eqref{xi=0block2}. So we can only obtain an inversion formula for the  building blocks, which means that the poles and  residues correspond to the building blocks and their coefficients. In \eqref{qexpan} we know that the multiplets block can be written as a polynomial in $q$. This motivates us to write  $\mathcal{G}^0$ as a power series expansion
\begin{equation}
\mathcal{G}^0(x,q)=\sum_k q^kf_k(x),   \label{q}
\end{equation}
then  try to do harmonic analysis for every individual $f_k(x)$. %}{The previous boundary condition is generally not satisfied by the 4 point function, I delete the relevant discussion.}

To do the harmonic analysis, we need to define a Hilbert space and determine a set of orthogonal bases of it. The decomposition of $f_k(x)$ on these bases should be related to its block expansion (by a contour deformation as usual). The basic logic for defining this Hilbert space and finding its  orthogonal bases  is to find an operator, which should be hermitian when acting on the desired Hilbert space. Then its normalizable eigenfunctions consist of a complete set of bases of the Hilbert space. The inner product and the boundary conditions, which ensure the hermitian condition, as well as the normalizable condition, define the needed Hilbert space. In fact, as we will see later, the harmonic analysis here is a special case of the harmonic analysis in 1D CFT with non-identical external operators.

\subsection{Harmonic analysis for \secmath{\xi=0}}
%\yu{}{This section is rewritten.}
   
The natural operator we can use is $\mathcal{D}_k$ in \eqref{D_k}.
In fact, this operator has already appeared in the calculation of usual conformal blocks in 1D CFT. Recall that the Casimir equation for four non-identical external operators in 1D CFT is
\begin{equation}
    D_xf(x)=\Delta(\Delta-1)f(x) \label{Dxequ}
\end{equation}
where
\begin{equation}
    D_x=x^2(1-x)\partial_x^2-(1+a+b)x^2\partial_x-abx  \label{Dx}
\end{equation}
and
\begin{equation}
    a=-\frac{\Delta_{12}}{2},\qquad b=\frac{\Delta_{34}}{2}.
\end{equation}
It is easy to see that when $\Delta_{12}=0, \Delta_{34}=2k$ or $\Delta_{12}=-2k, \Delta_{34}=0$ it becomes $\mathcal{D}_k$ in \eqref{D_k}.
There are at least two methods known to find a complete bases of a Hilbert space which makes $\mathcal{D}_k$ hermitian: the harmonic analysis and the alpha space method \cite{Hogervorst:2017sfd}. We briefly introduce these two methods in the following.

\subsubsection*{Review of the harmonic analysis/alpha space}
Though the 1D harmonic analysis with identical external operators is well-known \cite{Maldacena:2016hyu}, there are few literatures treating the non-identical case systematically. On the other hand, the alpha space approach is based on the same Strum-Liouville problem but with different boundary conditions. Here we review some facts about these non-identical case and point out some subtleties.

We should consider a  Strum-Liouville problem with $D_x$ in \eqref{Dx}. Then
the measure reads
\begin{equation}
\mu(x)=\frac{|1-x|^{a+b}}{x^2},  \label{measure}
\end{equation}
and the inner product is
\begin{equation}
    (f,g)=\int dx\mu f^*g.  \label{inner}
\end{equation}
Note that in this general case, the integral domain for the harmonic analysis in \eqref{inner} is not restricted to $x\in(0,2)$ as in the identical case \cite{Maldacena:2016hyu}, which comes from the invariance under $1\leftrightarrow2$ or $3\leftrightarrow4$. On the other hand, the integral domain for the alpha space is $x\in(0,1)$.

The fundamental eigen-functions for $D_x$ have the forms 
\begin{equation}
\begin{aligned}
       G_{\Delta}(x)&=x^{\Delta}\ _2F_1(\Delta+a,\Delta+b,2\Delta;x),\\
    G_{1-\Delta}(x)&=x^{1-\Delta}\ _2F_1(1-\Delta+a, 1-\Delta+b,2-2\Delta;x).
\end{aligned}
\end{equation}
In the harmonic analysis, the complete bases are the standard (bosonic) conformal partial waves (CPW), which can be obtained by the shadow integral
\begin{equation}\label{shadowint}
    \Phi_{\Delta}^{\Delta_1,\Delta_2,\Delta_3,\Delta_4}(x)=\frac{1}{2}\int_{-\infty}^{\infty}dy\frac{|x|^{\Delta}}{|y|^{\Delta+\Delta_{12}}|x-y|^{\Delta-\Delta_{12}}|1-y|^{1-\Delta+\Delta_{34}}}. 
\end{equation}
These conformal partial waves are  linear combinations of the above two basic solutions, with different combinational coefficients in $x<1$ and $x>1$. The dimensions of the above CPWs are the ones of the so-called principal series and the discrete series  representations. The shadow integral is  convergent  when $\Delta, \Delta_1, \Delta_2, \Delta_3, \Delta_4$ take values in the principal series $\frac{1}{2}+ir$.

In  the alpha space approach \cite{Hogervorst:2017sfd}, one impose that the CPW'\footnote{We use CPW' to denote the complete bases in the alpha space approach.} should be finite at $x=1$.  The CPW' has the following form:
\begin{equation} \label{CPW'}
    \Phi_{\alpha}^{\Delta_1,\Delta_2,\Delta_3,\Delta_4}(x)=z^{-a} \ _2F_1\left( \frac{1}{2}+a+\alpha, \frac{1}{2}+a-\alpha, 1+a+b; -\frac{1-z}{z}\right)
\end{equation}
 where $\alpha\equiv\Delta-\frac{1}{2}$ and $ \Phi_{\alpha}^{\Delta_1,\Delta_2,\Delta_3,\Delta_4}(1)=1$. 
It can be written as:
 \begin{equation}
      \Phi_{\alpha}^{\Delta_1,\Delta_2,\Delta_3,\Delta_4}(x)=\frac{1}{2}\left(Q(\alpha)G_\alpha(x)+Q(-\alpha)G_{-\alpha}(x)\right)
 \end{equation}
 where $G_\alpha(x)$ is the 1D block:
 \begin{equation}
     G_\alpha(x)=z^{\alpha+\frac{1}{2}}\ _2F_1\left(\frac{1}{2}+a+\alpha, \frac{1}{2}+b+\alpha, 2\alpha+1; z\right)
 \end{equation}
and
\begin{equation}
    Q(\alpha)=\frac{2\Gamma(-\alpha)\Gamma(1+a+b)}{\Gamma(\frac{1}{2}+a-\alpha)\Gamma(\frac{1}{2}+b-\alpha)}.
\end{equation}
The dimensions $\Delta_i, i=1,\cdots 4$ of CPW' lie in the  principal series.

From the differential equation \eqref{Dxequ}, the solution has leading behavior\footnote{When $a+b=0$, there could be subleading logarithmic correction.} $\sim(x-1)^\nu$ as $x\to1$, with the exponent $\nu=-a-b$ or $\nu=0$. From this point of view, the alpha space boundary condition (at $x=1$) just picks the solution with $\nu=0$: for $a+b>0$, CPW' in \eqref{CPW'} is the unique solution with exponent $\nu=0$; for $a+b=0$, which is degenerate and corresponds to the identical external operator case, requiring the finiteness near $x=1$ rules out the logarithmic behaviour $\sim\text{log}|1-x|$ and  picks out a unique solution \eqref{CPW'}; for $a+b<0$, one can obtain the CPW' by analytic continuation from $a+b>0$. On the other hand, the CPW \eqref{shadowint} in harmonic analysis is a solution with $\nu=-a-b$, so it is different from CPW'.

Notice that the measure \eqref{measure} near $x=1$ behave as $|1-x|^{a+b}$. Thus when  calculating the norm of CPW or CPW', one generically finds divergent terms in the integrand near $x=1$ when the absolute value of $a+b=2(\Delta_{34}-\Delta_{12})$ becomes large. The difference is that the norm of CPW diverges when $a+b$ is positive and large, while the norm of CPW' diverges when $a+b$ is negative and large.

\subsubsection*{The $\xi=0$ case}

For the $\xi=0$ inversion formula, we can use either CPW or CPW'. 
In our case, we prefer the alpha space approach because $a+b=k\in\mathbb{N}^{\geq 0}$ in \eqref{D_k}, at which the CPW' is always normalizable.

More precisely, 
the shadow integral formula in harmonic analysis start with the condition that  the internal operator as well as all the external operators needs to lie in the principal series such that the  shadow integral is convergent \cite{Simmons-Duffin:2017nub,Karateev:2018oml}. For real external dimensions, the shadow integral is generically divergent. A standard way to deal with this case is through analytic continuation from the case of principal external dimension. 
But there are subtleties when doing this continuation: the poles in the integrand may cross the contour of integration for $\Delta$, as shown in \cite{Simmons-Duffin:2017nub} for the $d>1$ case, and similarly for the $d=1$ case here. From another point of view,  for non-identical external operators with physical (real) dimensions,  when $a+b$ becomes large and positive,  CPW could be non-normalizable and the Casimir operators stop being self-adjoint due to the divergences at $x=1$.  It could be
possible that in this case the inner product can simply be modified by defining the integrals
by subtracting divergences near $x=1$. 
Though one can modify the harmonic analysis to apply in the case when $a+b=k\geq1$, we choose to use alpha space approach which in the  case at hand does not have this complexity. In the alpha space approach, the CPW' is normalizable and the Casimir is self-adjoint without any modification of the inner product.

\subsection{The inversion formula}

Now we can give out our inversion formula. First, look at the singular behaviour in the inversion function \eqref{inversionfun1}. If there are poles at $\xi=\xi_l=0$ and $\Delta=\Delta_m$, then there are $\xi=0$ operators in the propagating channel. Such a multiple pole  corresponds to the following block:
\begin{equation}
    \frac{1}{\xi^r}\frac{1}{\Delta-\Delta_m} \Longleftrightarrow \frac{x^{\Delta_m}(1+\sqrt{1-x})^{2-2\Delta_m} }{\sqrt{1-x}}\left(\frac{y}{x\sqrt{1-x}}\right)^r.
\end{equation}
Combined with the coefficients, we can obtain the contribution of all the $\xi=0$ operators, which is denoted as $\mathcal{G}_0$. We may expand $\mathcal{G}_0$ as in \eqref{q}, and for every $f_k(x)$ we define the following inversion function
\begin{equation}
    I_k(\Delta)=(f_k,\Phi_{k,\Delta}),
\end{equation}
where  $\Phi_{k,\Delta}(x):=\Phi^{\Delta_1,\Delta_2,\Delta_3,\Delta_4}_\alpha(x)$ with $-\Delta_{12}=0, \Delta_{34}=k$ and the inner product is defined in \eqref{inner}. The next step is to find the poles and the corresponding residues of every inversion function $I_k(\Delta)$, altogether of which give the decomposition of $\mathcal{G}_0$ in terms of the building blocks.

Note that in this case the (product of) OPE coefficients can not be read off directly from the residue of the inversion function. This is because we can only get the building block expansion of $\mathcal{G}_0$, while the full $\xi=0$ multiplet block is a linear combination of the building blocks. To calculate the OPE we need an extra step connecting these two expansions of $\mathcal{G}_0$. 
In the following subsections, we will show an example in BMS free scalar theory to verify our calculation of $\xi=0$ GCA block expansion. There we  will show how to obtain the OPE from the building block expansion.

Finally, we comment that there is a clever way\footnote{At least for our example of BMS free scalar.} to read the poles and residues of the inversion functions. Strictly speaking, one needs to substitute the full  CPW' in the inner product \eqref{inner} with $f_k(x)$. As a result, one will find un-physical shadow poles in the inversion function besides physical poles. This is the same as the inversion formula based on the harmonic analysis in higher dimension \cite{Caron-Huot:2017vep}.  However, we know that  the  CPW' are the linear  combinations of fundamental solutions $w^{(1)}_{k,\Delta}$ and $w^{(2)}_{k,\Delta}$. If one decompose CPW' into these two solutions from the beginning and follow the step to get the inversion formula, then effectively one can substitute only $w^{(1)}_{k,\Delta}$ into the inner product and the final result will only contain physical poles\footnote{The residue is also correct because the combination coefficients of $w^{(1)}_{k,\Delta}$ and $w^{(2)}_{k,\Delta}$ are  canceled precisely by the normalization factor.}. We will use this simple fact in the discussion of the inversion function in BMS free scalar.

\section{\secmath{\xi=0} multiplet in BMS free scalar model} \label{example sec}

In this section, we discuss $\xi=0$ multiplets in the BMS free scalar. The BMS$_3$ algebra is isomorphic to GCA$_2$ \cite{Bagchi:2012cy}, so our analysis can applies to the field theories with BMS symmetry. Here we focus on the  free BMS scalar model constructed in \cite{Hao:2021urq}, which provides an concrete example to see the block expansion of 4-point function in terms of $\xi=0$ multiplets.

\subsection{Review of the BMS free scalar}
The action of a BMS-invariant free scalar on a cylinder parameterize by$(\sigma,\tau)$ with $\sigma\sim \sigma+2\pi$ reads
\begin{equation}S = \frac{1}{4\pi} \int d\sigma d\tau \left( \partial_{\tau} \phi \right)^2.
\end{equation}
Having chosen the highest weight vacuum, the two-point function of the fundamental fields $\phi$ on the plane $(x,y)$ is
\begin{eqnarray}
\langle\phi(x_1,y_1)\phi(x_2,y_2)\rangle 
  = -\frac{y_1-y_2}{x_1-x_2}.
\end{eqnarray}

 There exist two kinds of primary operators in this theory: one kind includes two operators involving the derivatives on the fundamental fields, $O_0$ and $O_1$, which are defined as
\begin{equation}
O_0(x,y)\equiv i\partial_y \phi(x,y), \hs{3ex} O_1(x,y)\equiv i\partial_x \phi(x,y)\label{dphi},
\end{equation}
and the other kind consists of  the vertex operators
\begin{equation}\label{valpha}
V_\alpha (x,y)\equiv:e^{\alpha \phi(x,y)}:.
\end{equation}
Remarkably, $\mathbf{O}=(O_0,O_1)$ is a rank-2 primary multiplet, with conformal dimension and boost charge
\begin{equation}
\mathbf{\Delta}=\left(\begin{matrix}
   1 &0  \\
   0& 1\end{matrix}\right),\hs{3ex}
  \boldsymbol{\xi}=\left(\begin{matrix}
   0&0  \\
   1&0\end{matrix}\right).
\end{equation}
On the other hand, the vertex operator $V_{\alpha},\alpha\in \mathbb{R}$ or $i\mathbb{R}$ is a singlet primary operator with $\Delta=0,\xi=-\alpha^2/2$. Note that $\mathbf{O}$ is a primary multiplet, and automatically a quasi-primary multiplet. There is a class of  quasi-primary multiplets which are the composite operators with multiple $\mathbf{O}$. They are the descendant operators in the vacuum family as well as the $\mathbf{O}$ family so that they are all with vanishing boost charge $\xi$. Putting the vacuum module and the $\mathbf{O}$ module together, we summarize, up to $\Delta=3$, the number of states, quasi-primary states, primary states  and also the organization of the multiplets in the table below. In the last line, we use numbers in bold font to indicate the rank of the multiplets. For example, ${\bf 3}+{\bf 2}$ means that the 5 states with $\Delta=2$ split into a multiplet of rank $3$ and a multiplet of rank $2$.
\begin{figure}[h]\begin{center}
    \begin{tabular}{|c|c|c|c|c|}
\hline
conformal weight  &  $\Delta=0$ &  $\Delta=1$ & $\Delta= 2$ & $\Delta= 3$  \\
\hline
 \# of states &  1 &  2 &  5 &  10  \\
\hline
 \# of quasi-primaries  &  1 &  2 &  3 &  4  \\
\hline
 \# of primaries &  1 &  2 &  0 &  0 \\
\hline
multiplets &  {\bf 1} &  {\bf 2} &  {\bf 3}+{\bf 2} &  {\bf 3}+{\bf 1}+{\bf 4}+{\bf 2}\\
\hline
\end{tabular}\\
\end{center}
\caption{
States up to $\Delta=3$
}
\end{figure}
We want to emphasize here that with weight $\Delta$, there are $(\Delta+1)$ quasi-primary operators, forming a rank-$(\Delta+1)$ multiplet. The other multiplets with the same weight $\Delta$ are actually constructed from the descendants of the quasi-primaries with weights smaller than $\Delta$. 

The results of correlation functions for these primary operators are as follows. The two-point functions of $O_i, (i=1,2)$ are
\begin{equation}
\begin{aligned}
     \langle O_0(x_1,y_1)O_0(x_2,y_2) \rangle&=0,
\\ \langle O_0(x_1,y_1)O_1(x_2,y_2) \rangle&=\frac{1}{x_{12}^2},
\\
\langle O_1(x_1,y_1)O_1(x_2,y_2) \rangle&=-\frac{2y_{12}}{x_{12}^3}
\end{aligned}
\end{equation}
where $x_{12}=x_1-x_2$, $y_{12}=y_1-y_2$. The two-point functions above agree with the general result for a $\xi=0$ rank-2 multiplet.
All three-point functions within the multiplet vanish, namely,
\begin{equation}
    \langle O_a(x_1,y_1)\,O_b(x_2,y_2) \, O_c(x_3,y_3)\rangle=0, \qquad a,b,c=1,2.
\end{equation}
Next, the vertex operator $V_\alpha$ in \eqref{valpha} satisfies the following OPE
\begin{equation}\label{vvope}
V_\alpha(x',y') V_\beta(x,y) = e^{-\alpha \beta \frac{y'-y}{x'-x}} ~V_{\alpha+\beta}+\cdots,
\end{equation}
 which implies the two-point functions among them
\begin{equation}
\langle V_{\alpha}(x_1,y_1)V_{\beta}(x_2,y_2) \rangle= \begin{cases}
  e^{\alpha^2\frac{y_1-y_2}{x_1-x_2}}, &  \alpha+\beta=0, \\
  0, & \alpha+\beta\neq0.
\end{cases}  
\end{equation}
More generally the nonvanishing correlation functions of the vertex operators have the following form
\begin{equation}
    \langle \prod_{k=1}^{n}V_{\alpha_k}(x_k,y_k)\rangle=\text{exp}\left\{\sum_{i<j}^n(-\alpha_i\alpha_j)\frac{y_i-y_j}{x_i-x_j}\right\}
\end{equation}
with the condition
\begin{equation}
    \sum_k\alpha_k=0.
\end{equation}
Finally, let us consider the OPE between the multiplet and the vertex operators,
\begin{equation}
\begin{aligned}
O_0(x',y')V_\alpha(x,y)&= - \frac{i\alpha}{x'-x} V_\alpha(x,y),\\
O_1(x',y')V_\alpha(x,y)&=  \frac{i\alpha(y'-y)}{(x'-x)^2} V_\alpha(x,y),
\end{aligned}
\end{equation}
which means that the two-point functions between them always vanish. 

We are interested in the OPE of $V_\alpha V_{-\alpha}$, 
\begin{equation}
    V_\alpha V_{-\alpha}\sim \{\mathbf{1}\}+\{(O_0,O_1)^\mathrm{T}\}+\{(M,T,K)^\mathrm{T}\}+...
\end{equation}
where on the right-hand side we organize the multiplets according to their conformal weights. The first term is the identity singlet, the second term is $(O_0,O_1)^\mathrm{T}$ multiplet which is of rank 2, the third term is the rank-3 stress tensor multiplet $\mathbf{T}=(M,T,K)^\mathrm{T}$. One remarkable fact is that all the quasiprimary operators on the right-hand side must have vanishing charge. 
We want to check the contributions from these three multiplets. The OPE coefficient of the identity is 1. For $(O_0,O_1)^\mathrm{T}$,
using the Wick theorem, we find the non-vanishing three-point functions are
\begin{equation}
\begin{aligned}
\langle O_0(x_1,y_1)V_\alpha(x_2,y_2)V_{-\alpha}(x_3,y_3)\rangle &=\frac{-i\alpha}{x_{12}}e^{\alpha^2\frac{y_{23}}{x_{23}}}+\frac{i\alpha}{x_{13}}e^{\alpha^2\frac{y_{23}}{x_{23}}},\\
\langle O_1(x_1,y_1)V_\alpha(x_2,y_2)V_{-\alpha}(x_3,y_3)\rangle &=\frac{i\alpha \,y_{12}}{x_{12}^2}e^{\alpha^2\frac{y_{23}}{x_{23}}}-\frac{i\alpha\, y_{13}}{x_{13}^2}e^{\alpha^2\frac{y_{23}}{x_{23}}}.
\end{aligned}
\end{equation}
This three-point function can be written as the general form of the 3-point function \eqref{3pointA} \eqref{3pointB}, with $c_0=i\alpha, c_1=0$. So the contribution from the rank-2 multiplet $\mathbf{O}$ to the stripped four-point function is
\begin{equation}
g_{\mathbf{O}}=2c_0c_1g_0+c_0^2g_1=-\alpha^2g_1 \label{rank2},
\end{equation}
where
\begin{equation}
\begin{aligned}
g_0&=x^\Delta\ _2F_1(\Delta,\Delta,2\Delta;x)=x\ _2F_1(1,1,2;x),\\
g_1&=qx^\Delta\ _2F_1(\Delta,\Delta+1,2\Delta;x)=qx\ _2F_1(1,2,2;x).
\end{aligned}
\end{equation}
For the stress tensor rank-3 multiplet $(M,T,K)^\mathbf{T}$, it has three quasi-primaries
\begin{equation}
    T(x,y)=-:\partial_x\phi\partial_y\phi: ,\quad M(x,y)=-:\partial_y\phi\partial_y\phi: , \quad K(x,y)=-\frac{1}{2}:\partial_x\phi\partial_x\phi:,
\end{equation}
with weight and boost charge
\begin{equation}
\mathbf{\Delta}=\left(\begin{matrix}
   2 &0 &0 \\
   0 &2 &0 \\
   0 &0 &2\end{matrix}\right),\hs{3ex}
  \boldsymbol{\xi}=\left(\begin{matrix}
   0&0&0  \\
   1&0&0\\
   0&1&0 \end{matrix}\right).
\end{equation}
By the Wick theorem, we work out the three point function
\begin{equation}
\langle M(x_1,y_1)V_\alpha(x_2,y_2)V_{-\alpha}(x_3,y_3)\rangle =-\alpha^2(\frac{1}{x_{12}^2}+\frac{1}{x_{13}^2}-\frac{2}{x_{12}x_{13}})e^{\alpha^2\frac{y_{23}}{x_{23}}}.
\end{equation}
It can be written as the general form \eqref{3pointA} with $c_0=-\alpha^2$. Similarly, the three-point functions $\langle TV_\alpha V_{-\alpha}\rangle$ and $\langle KV_\alpha V_{-\alpha}\rangle$ determine the coefficients $c_1=c_2=0$.  So the contribution from the rank-3 multiplet $\mathbf{T}$ is:
\begin{equation}
g_{\mathbf{T}}=\frac{1}{2}c_0^2g_2=\frac{\alpha^4}{2}g_2 \label{rank3}
\end{equation}
where
\begin{equation}
g_2=q^2x^\Delta\ _2F_1(\Delta,\Delta+2,2\Delta;x)=q^2x^2\ _2F_1(2,4,4;x).
\end{equation}

\subsection{Four-point function of vertex operators}
To check our discussion in section 4, we consider the global block expansion of the following four-point function
\begin{equation}
\langle V_{\alpha}(x_1,y_1)V_{-\alpha}(x_2,y_2) V_{\alpha}(x_3,y_3)V_{-\alpha}(x_4,y_4) \rangle= e^{\alpha^2\frac{y_{12}}{x_{12}}}e^{-\alpha^2\frac{y_{13}}{x_{13}}}e^{\alpha^2\frac{y_{14}}{x_{14}}}e^{\alpha^2\frac{y_{23}}{x_{23}}}e^{-\alpha^2\frac{y_{24}}{x_{24}}}e^{\alpha^2\frac{y_{34}}{x_{34}}}
\end{equation}
Note that this is a four-point function of the singlets having the same weights and charges, where our general analysis is valid.
To go further, we consider the stripped four-point function with $s$-channel-contribution subtracted
\begin{equation}
G(x,q)=e^{\frac{\alpha^2qx}{x-1}}=\sum_{k=0}^\infty q^kG_k(x),
\end{equation}
where
\begin{equation}
G_k=\frac{\alpha^{2k}x^k}{k!(x-1)^k}  \label{G_k}.
\end{equation}
Note that we have the fact in mind that the contributions come from the $\xi=0$ multiplets.
Before using the inversion formula, we can expand it order by order as follows. Comparing with the contribution from the multiplet of rank $k$ \eqref{block462},
\begin{equation}
G_k=\sum_{\Delta}A_{k,\Delta} x^\Delta\ _2F_1(\Delta,\Delta+k,2\Delta;x)  \label{G_kexpan}
\end{equation}
we can check the consistency of the picture. For $k=0$,
\begin{equation}
G_0=1     \label{iden}
\end{equation}
which is the contribution from the identity operator. We have
\begin{equation}
A_{0,0}=1,\ \ A_{0,\Delta}=0\ \ (\Delta\neq0).
\end{equation}
For $k=1$,
\begin{equation}
  G_1=\frac{\alpha^2x}{x-1}.
\end{equation}
Notice that the block reads
\begin{equation}
x\ _2F_1(1,1+1,2;x)=-\frac{x}{x-1},
\end{equation}
so from \eqref{G_kexpan}, we know
\begin{equation}
A_{1,1}=-\alpha^2,\ \ A_{1,\Delta}=0\ \ (\Delta\neq1)
\end{equation}
This matches with \eqref{rank2}, the contribution from the multiplet of rank $2$. For $k=2$,
\begin{equation}
G_2=\frac{\alpha^4x^2}{2(x-1)^2}=\frac{\alpha^4}{2}x^2\ _2F_1(2,2+2,4;x),
\end{equation}
leads to
\begin{equation}
A_{2,2}=\frac{\alpha^4}{2},\ \ A_{2,\Delta}=0\ \ (\Delta\neq2),
\end{equation}
which agrees with \eqref{rank3}, the contribution from the multiplet of rank $3$. Generally, for $G_k$, notice that
\begin{equation}
g_k(\Delta=k)\equiv x^k\ _2F_1(k,k+k,2k;x)=\frac{x^k}{(1-x)^k}.
\end{equation}
Comparing it with \eqref{G_kexpan} and \eqref{G_k}, we conclude
\begin{equation}
A_{k,\Delta}=\delta_{k,\Delta}\frac{(-1)^k \alpha^{2k}}{k!}. \label{Acoe}
\end{equation}
From the general form \eqref{block462} and \eqref{Acoe}, we solve the three-point coefficients
\begin{equation}
c_{i,k}=\delta_{0i}\sqrt{(-1)^k}\alpha^k,\ \ k\in N.
\end{equation}
We conclude that the quasi-primary operators with weight $k$, charge $\xi=0$ appearing in the  $V_{a}V_{-a}$ OPE form a rank-$(k+1)$ multiplet. The three-point coefficient is the above $c_{i,k}$.

Another way to get the coefficients $A_k$ in a closed form is to consider the inversion formula discussed in section \ref{inversionsec}, from which we can calculate
\begin{equation}
(w^{(1)}_{k,\Delta},G_k)=\frac{(-1)^{-k}\Gamma(1-k)\Gamma(k+\Delta)}{(\Delta-k)(\Delta+k-1)B(\Delta+k,\Delta-k)\Gamma(\Delta)}\frac{\alpha^{2k}}{k!}.
\end{equation}
Picking up the residues, we have the coefficients
\begin{equation}
\text{Res}|_{\Delta=1-k}(w^{(1)}_{k,\Delta},G_k)=\frac{(-1)^{-k}\alpha^{2k}}{k!},
\end{equation}
which is \eqref{Acoe}.

In general, one can read the block expansion of the arbitrary stripped four-point functions $G(x,q)$ in a systematic way order by order. The first step is to expand $G(x,q)$ in terms of $q^k$,
\begin{equation}
G(x,q)=\sum_{k=0}^{\infty} q^k g_k(x),
\end{equation}
where $g_k(x)$ depends on $x$ only. Then one should expand each $g_k(x)$ in terms of the $SL(2,R)$ block $\frac{1}{k!}x^Q\ _2F_1(Q,Q+k,2Q,x)$,
\begin{equation}
g_k(x)=\sum_{Q}A_{k,Q}\frac{1}{k!}x^Q\ _2F_1(Q,Q+k,2Q,x).
\end{equation}
The discussion in section \ref{inversionsec} will help us find the coefficients $A_{k,Q}$ in a closed form. Or, one can use the direct expansion in terms of the powers of $x$. Considering the expansion of $g_k(x)$ in $x$,
\begin{equation}
g_k(x)=\sum_{z} B_z x^z,
\end{equation}
and the expansion of $SL(2,R)$ block
\begin{equation}
\frac{1}{k!}x^Q\ _2F_1(Q,Q+k,2Q,x)=\sum_{z} C_{Q,z} x^z,
\end{equation}
where
\begin{equation}
C_{Q,z}=\frac{2^{2 Q-1} \Gamma
   \left(Q+\frac{1}{2}\right) \Gamma (z)
   \Gamma (k+z)}{\sqrt{\pi } \Gamma (k+1)
   \Gamma (k+Q) \Gamma (-Q+z+1) \Gamma (Q+z)},
\end{equation}
one can get the coefficients $A_{k,Q}$ by solving the linear system,
\begin{equation}
C_{Q,z}A_{k,Q}=B_{z}.
\end{equation}
In practice, one can find the lowest $x^z$ in the power law expansion of $g_{k}(x)$, from which one can get the first $A_{k,Q_1}$. Then subtracting the contribution related to $A_{k,Q_1}$ and repeating this procedure, one can calculate the $A_{k,Q}$s order by order in $Q$. Note that $A_{k,Q}$ contains the contribution from different quasi-primary modules. Compared to the block expansion,
\begin{equation}
A_{k,Q}=\sum_{\Delta}\sum_{a=0}^{[\frac{r_\Delta-1}{2}]}\sum_{j=a}^{r_\Delta-1-a}c_{\Delta,i}c_{\Delta,j}f_a,
\end{equation}
where $Q=\Delta+a$, $i=r_\Delta-1-k-j$, and $r_\Delta$ is the rank of the quasi-primary operator with weight $\Delta$. For a certain propagating module, it will contribution to $A_{k,Q}$ with different $k$ and $Q$, with $0\leq k\leq r_{\Delta}-1$ and $\Delta \leq Q\leq \Delta+[\frac{r_\Delta-1}{2}]$. The first condition can be seen explicitly from \eqref{xi=0block1}. The second condition gives the infimum of $Q$, which is due to the fact that in the $a=0$ sector all $c_{\Delta,i}$s appear, while $\Delta+[\frac{r_\Delta-1}{2}]$ is the upper bounded due to the block expansion, but which may not be saturated, determined by the details of three-point coefficients $c_{\Delta,i}$. We propose the following prescription to read out the three-point coefficients $c_{\Delta,i}$ order by order. One first consider the $A_{k,Q}$s with the smallest $Q$, denoted as $Q_1$, which is equal to the conformal weight $\Delta$ of the concerned operator. Then consider all the $A_{k,Q_1}$ with different $k$, which will determine the rank $r_\Delta$,
\begin{equation}
A_{k,\Delta}=\sum_{j=0}^{r_\Delta-1}c_{\Delta,i}c_{\Delta,j}f_{a=0}|_{i=r_\Delta-1-k-j},
\end{equation}
where $f_{a=0}=1$. This is the same as the procedure to solve $c_i$ from $P_{k,\Delta}$ in \cite{Chen:2020vvn}. This procedure determines $c_{\Delta,i}$. There is no new information one can get from the $a\geq 0$ sector. Instead, one should subtract this contribution to get the data for other propagating quasi-primary multiplets. Next one consider the $A_{k,Q}$s with the second smallest $Q$, denoted as $Q_2$. If $Q_2-Q_1<1$, one should repeat the analysis for $Q_1$. Otherwise, one should subtract the contribution $\Delta=Q_1,a\geq 1$ from all $A_{k,Q_2}$ and repeat the discussion. Recursively, the data for higher $\Delta$ can be obtained in this way order by order.

\section{Conclusion and Discussions}

In this work, we extended our previous study on the bootstrap program of two dimensional  Galilean conformal field theory \cite{Chen:2020vvn} to the sector of  vanishing boost charge. Due to the presence of the null states in the $\xi=0$ sector, there are some novel features in the study.
Firstly, we showed that the existence of null states in the $\xi=0$ sector is a generic phenomenon in both the singlets and multiples. We analyzed the constraints on the theory coming from these null states by inserting the null states into the correlation functions.  We found that  there exist specific  fusion rules involving $\xi=0$ operators, as shown in \eqref{fusion0}.

Next, we computed the $\xi=0$ GCA multiplet block. In  \cite{Chen:2020vvn}, we obtained the $\xi\neq0$ multiplet block, which is roughly a linear combination of derivatives of the singlet block. In the  $\xi=0$ case this is no longer true because we need to mod out the null states to get the GCA block. We calculated the full GCA multiplet block by inserting the complete bases in the stripped four point function. This block can not be obtained by the Casimir equation. Nevertheless, we find the $\xi=0$ multiplet block can be written as a sum of building blocks, which can be obtained by the $SL(2,\mathbb{R})$ Casimir equation. %\tba

%In turns out that the $SL(2,\mathbb{R})$ Casimir equation is a special case of the one in 1D CFT with non-identical external operators. 
Furthermore, we tried to find the inversion formula  by including $\xi=0$ operators in the propagating channel. This required us to do harmonic analysis on the Galilean conformal algebra in the $\xi=0$ sector. Because the approach in \cite{Chen:2020vvn,Simmons-Duffin:2017nub} would lead to divergent terms destroying the normalizable and Hermitian conditions, we instead worked in the alpha space approach to find the inversion formula.  % In the $\xi=0$ sector we  borrow the alpha space method to get an inversion formula, since the 1D harmonic analysis in our case have practical complicity.
%\tba

As a consistent check of our study on the $\xi=0$ sector, we discussed the four-point functions of certain vertex operators in the BMS free scalar theory. The $\xi=0$ sector is the only sector appearing in the propagating channel. The fusion rule in this case is consistent with the one we found using the null states. By studying the global block expansion of the four-point function, we  reproduce the correct OPE by using a direct matching as well as the inversion formula. This provides nontrivial check on our formalism.% we obtained above.\tba

%(Future directions) \tba
In establishing the inversion formula for $\xi=0$ sector, we used the alpha-space approach.  In \cite{Chen:2022cpx}, the shadow formalism for 2D Galilean CFT was developed, but mainly focusing on $\xi\neq 0$ case. It would be interesting to further develop the shadow formalism in the $\xi=0$ sector. 

With the present work and the work in \cite{Chen:2020vvn}, we have established the framework to do bootstrap using the global GCA. We have been focusing on the analytic aspects of Galilean conformal bootstrap, studying the structure of the Hilbert space, computing the global blocks for the singlets and multiplets, doing harmonic analysis and setting up  the inversion formula. As the next step, it would be worth pursuing the numeric aspects of Galilean conformal bootstrap. At first thought, the bootstrap program could run into troubles as the theory is not unitary. But the surprising success in studying four-point functions in the generalized Galilean free field theory and BMS free scalar theory suggest that the real situation could be much better than we naively expected. It is certainly important to investigate this problem with more efforts.

\section*{Acknowledgments}
We are grateful to Y.F. Zheng for many valuable discussions. We would like to thank L. Apolo, B. Czech, C. Chang, W. Lai, J. Lu, W. Song, J. Wu, Z. Xiao, X. Xie, G. Yang, W. Yang, Y. Zhong for their stimulating questions and discussions in the workshops and seminars. The work is in part supported by NSFC Grant No. 11735001.
% The work is in part supported by NSFC Grant No. 11335012, No. 11325522 and No. 11735001.

\vspace{1cm}
\appendix
\renewcommand{\appendixname}{Appendix~\Alph{section}}

\begin{refcontext}[sorting=none]
\printbibliography
\end{refcontext}

\end{document}